\documentclass{aastex631}

\newcommand{\eg}[1]{{\color{red}{#1}}}

\usepackage{bm}
\usepackage{subfigure}
\usepackage{amsmath}
\usepackage{cases}
\usepackage{comment}
\usepackage{soul}
\usepackage{rotating}

\DeclareUnicodeCharacter{2212}{-}
\usepackage{newunicodechar}

\newunicodechar{≤}{\ensuremath{\leq}}
\usepackage{xcolor}
\usepackage{amssymb}
\usepackage{appendix}
\shorttitle{Polarizaton of Cassiopeia A}
\shortauthors{Mercuri et al.}
\usepackage{graphicx} 
\graphicspath{{./}{figures/}}

\begin{document}

\title{Revisiting X-ray polarization of the shell of Cassiopeia A using spectropolarimetric analysis}

\author{A. Mercuri}
\affiliation{Università della Calabria, via P. Bucci, cubo 33C, 87036, Rende (CS), Italy}

\author{E. Greco}
\affiliation{INAF - Osservatorio Astronomico di Palermo, Piazza del Parlamento 1, 90134 Palermo, Italy}
\affiliation{Anton Pannekoek Institute for Astronomy \& GRAPPA, University of Amsterdam, Science Park 904, 1098 XH Amsterdam, The Netherlands}

\author{J. Vink}
\affiliation{Anton Pannekoek Institute for Astronomy \& GRAPPA, University of Amsterdam, Science Park 904, 1098 XH Amsterdam, The Netherlands}

\author{R. Ferrazzoli}
\affiliation{INAF - Istituto di Astrofisica e Planetologia Spaziali, Via del Fosso del Cavaliere 100, 00133 Roma, Italy}

\author{S. Perri}
\affiliation{Università della Calabria, via P. Bucci, cubo 33C, 87036, Rende (CS), Italy}

\begin{abstract}
X-ray synchrotron radiation is expected to be highly polarized. Thanks to the Imaging X-ray Polarimetry Explorer (IXPE), it is now possible to evaluate the degree of X-ray polarization in sources such as supernova remnants (SNRs). Jointly using IXPE data and high-resolution Chandra observations, we perform a spatially resolved spectropolarimetric analysis of SNR Cassiopeia A (Cas A). We focus in the 3–6 keV energy band on regions near the shell dominated by nonthermal synchrotron emission. By combining IXPE’s polarization sensitivity with Chandra’s higher spatial and spectral resolution, we constrain the local polarization degree (PD) and polarization angle (PA) across the remnant. 
Our analysis reveals PD values ranging locally from 10\% to 26\%, showing significant regional variations that underscore the complex magnetic field morphology of Cas A. The polarization vectors indicate a predominantly radial magnetic field, consistent with previous studies. Thanks to the improved modeling of thermal contamination using Chandra data, we retrieve higher PD values compared to earlier IXPE analysis and more significant detections with respect to the standard IXPEOBSSIM analysis. Finally, we also estimate the degree of magnetic turbulence $\eta$ from the measured photon index and PD, under the assumption of an isotropic fluctuating field across the shell of Cas A. 

\end{abstract}




\section{Introduction}
Supernova remnants (SNRs) are the diffuse leftover of supernova (SN) explosions, the violent deaths of massive stars or white dwarfs. Shock fronts in SNRs are considered the most likely acceleration sites of Galactic cosmic rays (CRs), high-energy particles that can reach energies up to hundreds of TeV, through the so-called Diffusive Shock Acceleration (DSA) mechanism (e.g., \citealt{Drury1983}). In the DSA framework, a significant fraction ($\sim 10\%$) of the kinetic energy from the SN explosion is transferred to charged particles, which repeatedly cross the shock front back and forth, gaining energy at each crossing \citep{Amato2014}. Multi-wavelength spectral analyses have reinforced this hypothesis by revealing synchrotron radiation across the electromagnetic spectrum, from radio to X-ray frequencies. Radio emission is ascribed to electrons accelerated up to GeV energies, whereas the X-ray one arises from TeV electrons \citep{Vink2020}.
Thanks to X-ray spectroscopy, it is possible to study the interaction between SNRs and the surrounding interstellar/circumstellar medium (ISM/CSM). Indeed, as SNRs expand, they drive powerful shock waves that heat and compress the CSM/ISM, producing characteristic X-ray emission that can be broadly categorized into two types: thermal and nonthermal. The thermal emission, either due to shocked stellar fragments of the progenitor star (the ejecta) or to the swept-up ISM/CSM, arises from electrons and ions that have been heated to a Maxwellian distribution by the reverse or forward shock, respectively. On the other hand, the nonthermal emission is generated by relativistic electrons that are accelerated at the shock fronts and gyrate around the magnetic field lines emitting X-ray synchrotron radiation \citep{Slane1999, Parizot2006, Morlino2010, Vink2020}.

Synchrotron radiation is intrinsically polarized up to the 70\%–75\% level (depending on the slope of the electrons spectra - see equation (3.29) in \citet{GinzburgSyrovatskii1965}). However, the presence of magnetic turbulence can significantly decrease the polarization level of the radiation. Thus, the analysis of the polarization of radiation from SNRs is crucial for understanding the structure of the magnetic fields, including the ones present in the ISM and the CSM, and the degree of turbulence induced by the SNR expansion and its interaction with the surrounding medium. 
Observations of radio polarization have revealed that magnetic fields in young SNRs are often radially oriented \citep{Dickel1976}. In SN 1006 shell the magnetic field distribution has shown a predominantly radial configuration, with components almost aligned with the Galactic plane \citep{Reynoso2013}. The average polarization angle (PA) measured in the radio band for SN 1006 is approximately $-36.3^\circ \pm 0.4^\circ$.
Similarly, the Tycho's SNR is characterized by 
radio polarization degrees (PDs) ranging from 0 at the center to 7-8\% at the outer rim (\citealt{Kundu&Velusamy1971}; \citealt{Dickel1991}).

One of the most-studied and brightest SNR is Cassiopeia A (Cas A). Cas A is one of the youngest known SNRs in the Milky Way, being a 343 years old SN IIb-type remnant at a distance of 3.4 kpc \citep{Reed1995,Krause2008}, which shows many asymmetries and an overall clumpiness.  It is about 3 pc in radius and its shock front expands at a speed of 4000-6000 km/s \citep{Reed1995,Lawrence1995,Milisavljevic12013,Vink2022b, Sakai2024}.

Cas A shows a radially oriented magnetic field \citep{Rosenberg1970,Braun1987} and exhibits relatively low radio polarization, around 5\% close to the reverse shock and between 8\% and 10\% in the outer regions \citep{Anderson1995}. 
The predominantly radially oriented magnetic field in Cas A,
combined with the properties of synchrotron emission, predicts that the intrinsic polarization may differ significantly between the X-ray and radio bands. While X-ray synchrotron emission is confined to thin regions downstream of the shock, with a characteristic width given by $ l_{\text{loss}} \approx \Delta v \, t_{\text{loss}}$,
where $\Delta v = \frac{1}{4} V_{\text{sh}}$ is the downstream flow velocity, radio synchrotron emission originates from a much larger volume, with typical size of the order of $ l \gtrsim 10^{18} \, \text{cm} $, while the narrowness of the synchrotron filaments is typically $ \sim 5 \times 10^{16} \, \text{cm}$ \citep{Helder2012}.
 In the radio band, the longer path lengths result in significant depolarization due to variations in the magnetic field orientation along the line of sight. 
In contrast, the shorter path lengths probed by X-ray synchrotron emission minimize the impact of magnetic field variations allowing a higher PD to be observed. Furthermore, the steep slope of the X-ray synchrotron spectrum implies that the maximum intrinsic PD \( P_{\max} \) in the X-ray band is higher than in the radio \citep{GinzburgSyrovatskii1965,Bykov2009}.

 The Imaging X-ray Polarimetry Explorer (IXPE), resulting from a collaboration between NASA and the  Italian space agency (ASI), is the first telescope able to measure X-ray polarization \citep{Soffitta2021,Weisskopf2022}. It enables X-ray spatially resolved spectropolarimetric analysis for the first time, achieving an angular resolution of approximately $30''$ in the 2–8 keV energy band. One of the first studies made on extended sources by means of IXPE data was carried out by \citet{Vink2022}, who detected polarized X-ray emission in the 3–6 keV range from  Cas A. They found an overall PD for synchrotron radiation of about $2.5\%$, and nearly $5\%$ in the forward shock region, with a radially-oriented magnetic field, in agreement with radio observations \citep{Vink2022}. \\
Besides Cas A, IXPE observations revealed that also Tycho's SNR and the northeastern limb of SN1006 exhibit radially oriented magnetic fields, with average PD of $9 \pm 2\%$ over the entire Tycho remnant (\citealt{Ferrazzoli2023}) and of $22.4 \pm 3.5\%$ for the limb of SN1006 (\citealt{Zhou2023}).
In contrast, for the northwestern rims of SNR RX J1713.7-3946 and SNR Vela Jr (RX J0852.0−4622), a tangential magnetic field orientation in the acceleration regions was detected, with an average PD of $12.5 \pm 3.3\%$ and PD $= 16.4\% \pm 4.5\%$, respectively (\citealt{Ferrazzoli2024},\citealt{Prokhorov2024}). The latter result added a new example of tangential magnetic fields in young SNRs, challenging the previously observed dichotomy between radial magnetic fields in young SNRs and tangential fields in middle-aged ones.

In this study we perform a spatially resolved spectropolarimetric analysis of the SNR Cas A focusing on regions close to the shell, where the nonthermal synchrotron emission dominates. Given the IXPE limited spectral resolution, which is insufficient to disentangle detailed spectral features, such as continuum and emission lines, Chandra data are used to constrain the parameters of the thermal emission. Therefore, we couple the higher Chandra spatial and spectral resolution with the unique diagnostic potential on X-ray polarization provided by IXPE, in order to constrain the thermal and nonthermal emissions at the shock front. Then, we evaluate both the PD and the PA across the shell of Cas A in each selected region.
Moreover, we also perform a polarization analysis using the open-source software package \texttt{IXPEOBSSIM} \citep{Baldini2022}. From the derived PDs we estimate the degree of magnetic turbulence all around the Cas A front, within each selected region, to identify the most turbulent (low polarized) regions. The paper is organized as follows: in Section 2, we present the analysis methods, detailing the procedures used for the spectropolarimetric study and the analysis conducted with the IXPEOBSSIM software package. In Section 3, we report the results of these analyses, including both the spectropolarimetric findings and the outcomes of the IXPEOBSSIM-based analysis. In Section 4, we discuss the implications of our results and provide the main conclusions of the study. 


\section{Data analysis}
In order to extract the polarization properties of the emission around the entire shell of Cas A, we make use first of the Chandra high resolution data to deduce the typical parameters of both the thermal and non-thermal components of the emitted radiation. Then, the values obtained serve to constraint the best-fit of the IXPE Stokes parameters spectra. 

    \subsection{Chandra data}
\begin{figure}
    \centering
    \centerline{\includegraphics [width=15cm]{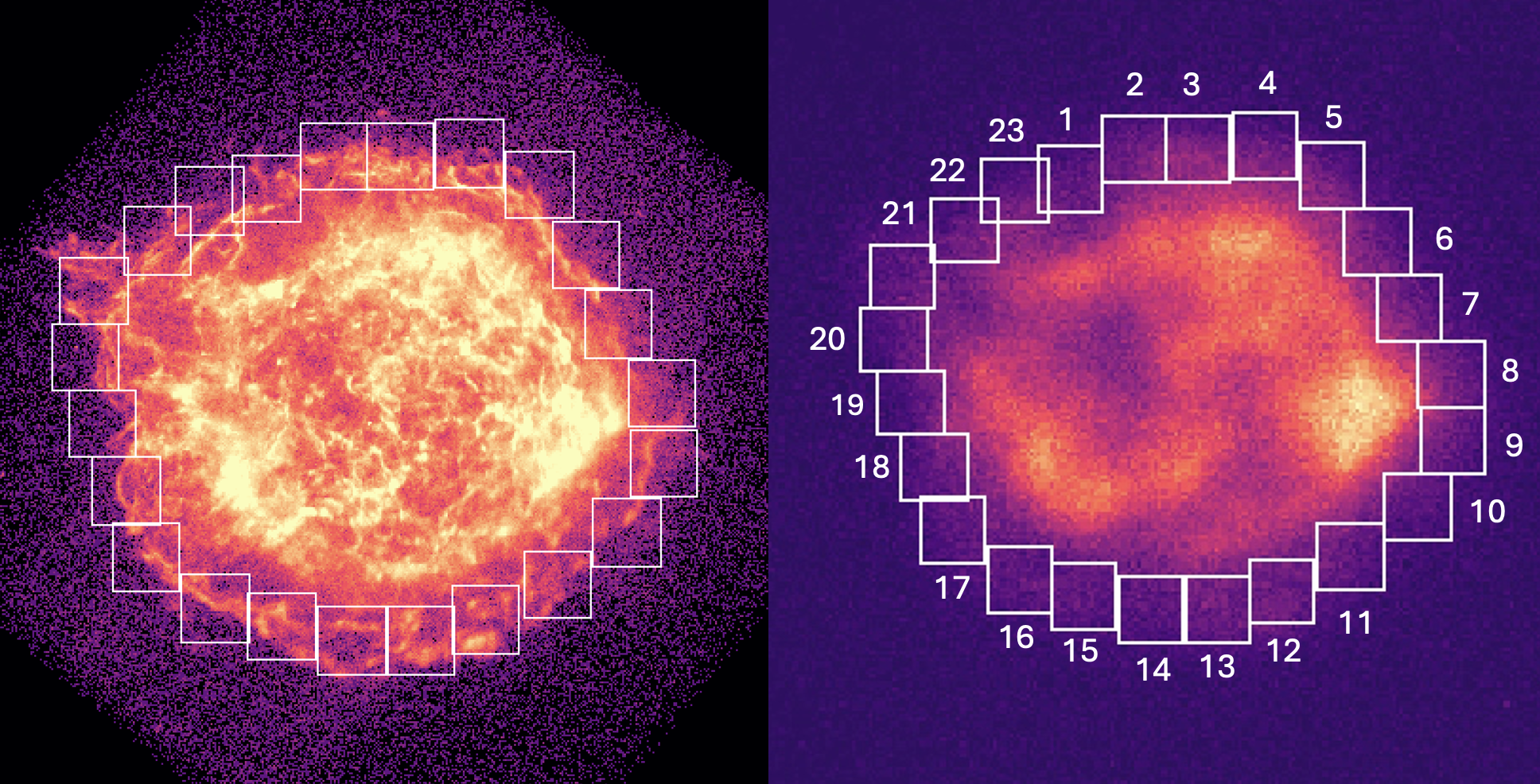}}
    \caption{Cas A images of Chandra (left panel), in the log scale with a pixel of size 0.98 arcsec, and IXPE (right panel), in the sqrt scale with a pixel of size 2.6 arcsec, in the 3-6 keV energy band. The regions selected for the spectral analysis around the SNR rims are indicated by white squares.}
    \label{obs}
\end{figure}

    We analyze the deepest single Chandra observation of Cas A (ObsID 4638; Exposure time: 164 ks, PI Hwang \citep{Hwang2004}). 
    The data are reprocessed with the CIAO v4.16 software using CALDB 4.11.0. We use the \texttt{chandra\_repro} task to reduce the observation. The image of Cas A, generated using the \texttt{fluximage} tool, is plotted on the left panel in Figure \ref{obs}. It is taken in the 3-6 keV energy band, in order to retain mainly the non-thermal contribution of the emission and to minimize the thermal contamination \citep{Vink2022}. Thus, we select 23 regions of size 42''x42'' across the whole shell of the remnant, as shown in the left panel in Fig. \ref{obs} where they are indicated by white squares. For each of the selected regions, we extract the spectra through the \texttt{specextract} tool, available within CIAO.
 For all the spectra we impose that each bin in energy must contain at least 25 counts, in order to use $\chi^2$ statistics; then, we subtract to each observed spectrum the background spectrum obtained from a nearby region immediately outside of the source. Spectral analysis is performed with PyXspec 2.1.2, the Python wrapper of XSPEC 12.13.1 \citep{Arnaud1996}.  One example of the obtained energy spectra is shown in Figure \ref{spettro_abb}.

\subsection{IXPE data}
 IXPE has observed Cas A from 2022 January 11 to January 29 for a total time of about 900 ks (obs. id. 01001301). 
The initial IXPE data release of the Cas A observations contains some inaccuracies in the sky coordinates due to phase-dependent effects. Specifically, the boom connecting the X-ray telescopes to the spacecraft experienced heating variations during its orbit around Earth, leading to periodic bending and shifts in the focal point within the detector coordinates. Despite the application of corrections, some of which are included in the publicly available level two event list, a residual pointing offset of approximately 2.5' persists.
    The event lists used \footnote{\url{https://zenodo.org/records/6597504}} have corrected WCS keywords, based on a cross correlation with Chandra observations \citep{Vink2022}.
    
    In addition, events likely to be particles rather than photons are removed based on their event tracks.
    Then, we carefully account for background contributions. They can be classified into two main types: the diffuse astrophysical X-ray background and the particle-induced background generated by the instrument. To further refine the separation between particle-induced events and photon events, we apply an energy-dependent background filtering technique following the method outlined in \citet{DiMarco2023} \footnote{\url{https://github.com/aledimarco/IXPE-background}}. This approach, which also utilizes event track information, leverages key differences in track morphology. X-ray tracks, produced by photoelectrons in the gas-pixel detectors, are compact with higher charge density, while particle-induced tracks tend to be more extended, straighter, and less dense. \citet{DiMarco2023} developed three rejection filters based on: (i) track size (\texttt{NUM\_PIX}), (ii) the energy fraction in the main cluster (\texttt{EVT\_FRA}), and (iii) the number of border pixels (\texttt{NUM\_TRK}). By applying energy-independent cuts-requiring \texttt{NUM\_PIX} $<$
    250, \texttt{EVT\_FRA} $>$ 0.82, and \texttt{NUM\_TRK} $<$ 2, we effectively eliminate the most likely particle-induced events.
    Despite this, a non-negligible background remains, so we select a region free of sources to perform a further background subtraction. 
    
    The squared regions along the bright rims selected in the Chandra observations are reported on the same location in the IXPE Cas A image (see the right panel in Figure \ref{obs}). The IXPE image of Cas A is taken in the 3-6 keV energy band, generated using the \texttt{xpselect} tool provided by \texttt{IXPEOBSSIM}. The Stokes parameters I, Q, and U spectra for all the 23 regions are extracted using \texttt{xselect}. \\ 
    We group the Stokes I spectra to contain at least 50 counts per bin, while we apply a constant 0.2 keV energy binning to the Stokes Q and U spectra, following the procedure described in \citet{Ferrazzoli2024}. \\

\subsection{Methods}

\begin{figure}
    \centering
    \centerline{\includegraphics[width=18cm]{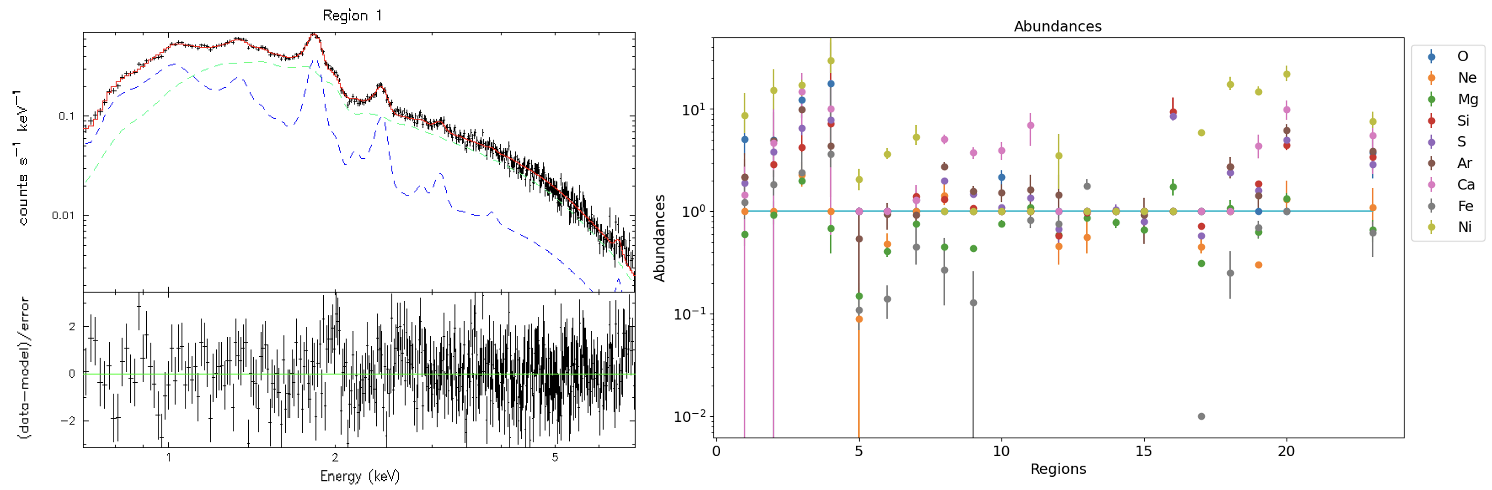}}
    \caption{\textit{Left panel}. Chandra emission spectrum of region 1. In the top panel, black points represent the observational data with their associated error bars. The solid red line shows the overall best-fit model, while the dashed blue line represents the thermal \texttt{vnei} component, and the dashed green line indicates the non-thermal \texttt{power-law} component. In the bottom panel, the residuals for each energy bin are displayed along with their corresponding errors. \textit{Right panel}. Chemical abundances derived from the best fit of Chandra data for the analyzed regions. The y-axis represents the abundances on a logarithmic scale. The abundances of regions 21 and 22 are not shown, as they are significantly affected by the Cas A jet.}
    \label{spettro_abb}
\end{figure}

The regions shape and size were chosen by finding the best balance between three different competing effects: i) each region has to be bigger than the IXPE point spread function to limit the bleeding of photons from close regions; ii) the bigger the regions the better the statistics, a crucial point especially to robustly constrain the polarization parameters iii) on the other hand, bigger regions are more likely to be affected by thermal contamination from the ejecta or to cause mixing between different polarization angles, hampering the diagnostic power on the polarization. We found the best trade-off by selecting 42''x42'' boxes along the bright synchrotron rims. In any case, as shown in Figure \ref{spettro_abb}, emission lines are still present in the spectra, indicating thermal contamination from either ejecta or ISM/CSM. 
To correctly take into account this additional thermal component, each Chandra spectrum is fitted using the model \texttt{TBabs(vnei+powerlaw)}, which includes a thermal, non-equilibrium ionization component \citep{Borkowski2001}, a powerlaw component to fit the synchrotron contribution, and the galactic absorption model \citep{Wilms2000}. As an example, the spectrum of region 1 fitted by the above model is displayed in Figure \ref{spettro_abb}, together with the residuals. We note here that addressing the exact origin of the detected thermal emission, i.e. ejecta/ISM/CSM, is beyond the scope of this paper, since we are only interested in model the observed thermal emission to accurately include its effects in the analysis of the IXPE observations. To achieve a good spectral fit with the minimum number of free parameters while ensuring an acceptable $\chi^2$ value relative to the degrees of freedom, an \texttt{F-test} is performed to determine which chemical abundance should be fixed and which should remain free to vary during the fitting process.
The best-fit parameters derived within all the 23 regions are reported in Table \ref{parameters}. In addition, we plot in Figure \ref{spettro_abb} the chemical abundances along with their error bars. Points without error bars correspond to abundances fixed to 1. For regions 4, 5, 6, and 23, where there is significant degeneracy between $\tau$, the ionization timescale,  and $kT$, the electron temperature, we calculate the parameter uncertainties by fixing one parameter while varying the other. Notice that regions 21 and 22 are not included in Table \ref{parameters} and in the plot in Figure \ref{spettro_abb}, since their abundances are too high, being regions very close to the Cas A jet. Errors are provided at the $90\%$ confidence level.
Almost all the regions show supersolar abundances, indicating the presence of shocked ISM/CSM and/or contamination from the shocked ejecta. This is particularly true for the northern side regions and the ones on the southeast side.

On the other hand, IXPE spectra are fitted with a model that combines Galactic absorption, an unpolarized ($PD$ is kept fixed to 0) thermal component, a non-thermal powerlaw component with PD and PA kept constant over the 3–6 keV energy band; in XSPEC syntax: \texttt{TBabs(polconst*vnei+polconst*powerlaw)}. 
Since the absorption can hardly be constrained with the $> 2$ keV photons, the values derived from the Chandra data analysis are used. For each region, the parameters of the thermal component are fixed to the best-fit values derived from the corresponding Chandra data fit.

\begin{table}
   \centering
   \tabcolsep=4pt
    \begin{tabular}{|c|c|c|c|c|c|c|c|c|}
\hline
Region & nH ($10^{22}$) & kT (keV) & $\tau$ ($10^{11} \mathrm{s}/\mathrm{cm^3}$)& $norm_{vnei}$ ($10^{-3} \frac{\mathrm{erg}}{\mathrm{cm^3s}}$)& $\Gamma$ & $norm_{pow}$ ($10^{-3} \frac{\mathrm{photons}}{\mathrm{keV cm^2 s}}$) & {\textbf{$\chi^2$}} & {\textbf{DOF}} \\
\hline
1 & 1.02$_{-0.07}^{+0.06}$ & 3.93$_{-0.84}^{+1.27}$ &  0.24$_{-0.02}^{+0.22}$  & 0.81$_{-0.28}^{+0.28}$ & 2.77$_{-0.08}^{+0.08}$ & 4.20$_{-0.14}^{+0.29}$ & 466 & 369 \\
\hline
2 & 1.13$_{-0.02}^{+0.04}$ & 1.15$_{-0.16}^{+0.16}$  & 0.51$_{-0.09}^{+0.14}$ & 1.73$_{-0.55}^{+0.56}$ & 2.80$_{-0.02}^{+0.03}$ & 6.25$_{-0.27}^{+0.33}$ & 462 & 331 \\
\hline
3 & 1.13$_{-0.01}^{+0.01}$ & 0.84$_{-0.02}^{+0.07}$ &  1.29$_{-0.18}^{+0.10}$ &  3.29$_{-1.05}^{+1.16}$ & 2.76$_{-0.07}^{+0.02}$ & 7.95$_{-1.14}^{+1.50}$ & 453 & 338 \\
\hline
4 & 1.02$_{-0.07}^{+0.02}$ & 1.76$_{-0.29}^{+0.08}$  & 0.44$_{-0.04}^{+0.02}$  & 0.41$_{-0.24}^{+0.29}$ & 2.71$_{-0.05}^{+0.05}$ & 4.78$_{-0.17}^{+0.15}$ & 367 & 329 \\
\hline
5 & 0.82$_{-0.05}^{+0.06}$ & 1.58$_{-0.14}^{+0.12}$  & 0.77$_{-0.10}^{+0.10}$  & 2.58$_{-0.12}^{+0.12}$ & 2.34$_{-0.07}^{+0.08}$ & 1.99$_{-0.20}^{+0.23}$ & 367 & 324 \\
\hline
6 & 0.98$_{-0.04}^{+0.04}$ & 2.44$_{-0.10}^{+0.13}$  & 0.527$_{-0.03}^{+0.04}$  & 2.71$_{-0.9}^{+0.14}$ & 2.72$_{-0.06}^{+0.06}$ & 2.99$_{-0.27}^{+0.22}$ & 402 & 323 \\
\hline
7 & 1.43$_{-0.03}^{+0.04}$ & 3.29$_{-0.27}^{+0.64}$ & 0.47$_{-0.05}^{+0.03}$ &  3.37$_{-0.72}^{+0.54}$ & \(>\)3.5 & 8.56$_{-0.59}^{+0.83}$ & 418 & 328 \\
\hline
8 & 1.32$_{-0.05}^{+0.07}$ & 1.04$_{-0.10}^{+0.10}$ & 3.29$_{-0.16}^{+0.10}$ & 10.60$_{-0.90}^{+0.90}$ & $<1.30$ & 0.11$_{-0.01}^{+0.11}$ & 363 & 271 \\
\hline
9 & 1.55$_{-0.14}^{+0.11}$ & 1.08$_{-0.12}^{+0.09}$ & 3.03$_{-0.85}^{+0.16}$ &  10.20$_{-1.40}^{+1.70}$ & $<1.30$ & 0.17$_{-0.01}^{+0.13}$ & 371 & 282 \\
\hline
10 & 1.82$_{-0.02}^{+0.02}$ & 0.87$_{-0.01}^{+0.02}$ &  \(>\)4 & 9.00$_{-0.34}^{+0.37}$ & $<1.30$ & 0.11$_{-0.01}^{+0.01}$ & 350 & 268 \\
\hline
11 & 1.82$_{-0.05}^{+0.05}$ & 0.87$_{-0.02}^{+0.05}$ &  2.49$_{-0.43}^{+0.51}$ &  9.22$_{-0.90}^{+0.73}$ & 2.60$_{-0.07}^{+0.07}$ & 0.24$_{-0.21}^{+0.22}$ & 338 & 298 \\
\hline
12 & 1.59$_{-0.11}^{+0.10}$ & 0.84$_{-0.09}^{+0.11}$ &  1.34$_{-0.37}^{+0.45}$ &  17.40$_{-5.60}^{+6.00}$ & 2.54$_{-0.27}^{+0.16}$ & 4.18$_{-1.61}^{+1.35}$ & 396 & 317 \\
\hline
13 & 1.81$_{-0.04}^{+0.07}$ & 0.81$_{-0.04}^{+0.02}$ &  2.57$_{-0.35}^{+0.31}$  & 8.70$_{-0.51}^{+0.60}$ & 2.66$_{-0.05}^{+0.04}$ & 5.42$_{-0.36}^{+0.29}$ & 344 & 319 \\
\hline
14 & 1.71$_{-0.03}^{+0.02}$ & 0.81$_{-0.03}^{+0.03}$ &  4.71$_{-1.04}^{+4.71}$ & 6.15$_{-0.51}^{+0.60}$ & 2.58$_{-0.03}^{+0.05}$ & 3.11$_{-0.24}^{+0.22}$ & 351 & 307 \\
\hline
15 & 1.99$_{-0.03}^{+0.02}$ & 0.84$_{-0.02}^{+0.01}$ & \(>\)4 & 10.10$_{-0.40}^{+0.40}$ & 2.29$_{-0.06}^{+0.05}$ & 2.78$_{-0.14}^{+0.15}$ & 428 & 316 \\
\hline
16 & 0.83$_{-0.05}^{+0.05}$ & 1.34$_{-0.12}^{+0.40}$ &  1.52$_{-0.12}^{+0.80}$  & 0.37$_{-0.07}^{+0.09}$ & 2.29$_{-0.06}^{+0.04}$ & 2.10$_{-0.10}^{+0.13}$ & 389 & 307 \\
\hline
17 & 1.09$_{-0.03}^{+0.09}$ & 1.31$_{-0.04}^{+0.08}$ &  1.61$_{-0.14}^{+0.27}$  & 8.11$_{-0.36}^{+0.70}$ & 1.37$_{-0.04}^{+0.11}$ & 0.29$_{-0.01}^{+0.03}$ & 449 & 321 \\
\hline
18 & 0.90$_{-0.02}^{+0.02}$ & 1.36$_{-0.04}^{+0.17}$ &  1.26$_{-0.13}^{+0.16}$  & 2.02$_{-0.14}^{+0.49}$ & 2.80$_{-0.02}^{+0.05}$ & 3.96$_{-0.26}^{+0.24}$ & 455 & 322 \\
\hline
19 & 0.80$_{-0.02}^{+0.03}$ & 1.66$_{-0.12}^{+0.22}$&  1.12$_{-0.20}^{+0.20}$  & 1.58$_{-0.15}^{+0.11}$ & 2.99$_{-0.06}^{+0.08}$ & 2.48$_{-0.18}^{+0.23}$ & 465 & 298 \\
\hline
20 & 0.83$_{-0.02}^{+0.03}$ & 1.42$_{-0.11}^{+0.23}$ &  1.48$_{-0.28}^{+0.31}$  & 1.05$_{-0.14}^{+0.15}$ & 2.98$_{-0.03}^{+0.05}$ & 3.07$_{-0.09}^{+0.07}$ & 429 & 300\\

\hline
23 & 0.88$_{-0.02}^{+0.05}$ & 3.78$_{-0.33}^{+1.15}$ & 0.31$_{-0.02}^{+0.03}$  & 0.34$_{-0.02}^{+0.03}$ & 2.47$_{-0.07}^{+0.07}$ & 2.17$_{-0.11}^{+0.19}$ & 422 & 354 \\
\hline
 \end{tabular}
\caption{Best-fit parameters obtained from the Chandra spectra. The parameters include the column density $nH$, the electron temperature $kT$,  the ionization timescale $\tau$, the normalization factor for the \texttt{vnei} model, the photon index $\Gamma$ and the normalization factor for the \texttt{powerlaw} model. In regions 4, 5, 6, and 23, due to strong degeneracy between $\tau$ and $kT$, uncertainties are estimated by holding one parameter fixed while varying the other. The parameters of regions 21 and 22 are not shown, as they are significantly affected by the Cas A jet. Uncertainties are at the 90\% confidence level.}
\label{parameters}
\end{table}

\subsubsection{Validation of the spectropolarimetric analysis}

This work represents the first spectropolarimetric analysis of a SNR using IXPE observations that includes a thermal component in the modeling. The goal is to measure the PD and PA through a spectropolarimetric analysis while accurately accounting for the thermal emission. This can be achieved only by understanding how the PD is influenced by the presence of a thermal component in the spectral fits and how the coupling between the polarization (\texttt{polconst}) and the \texttt{vnei} thermal model affects the results.

Given the novelty of our approach and the uncertainties on the use of the polarization component in presence of a significant thermal emission, we validated our methodology by generating synthetic IXPE spectra to be used as calibration sources. This is accomplished using the \texttt{fakeit} tool provided by the PyXspec software package, which allows us to simulate spectra derived from a user-defined model. The validation process is based on verifying that the fitting procedure, carried out using different possible physical models, applied to the synthetic spectra, returns the same parameter values given as input in the simulation. This approach is crucial to verify that when the thermal emission is not properly modeled when retrieving the polarization parameters, misleading results can be obtained; thus, we need to ensure that we can properly derive the physical characteristics of the plasma from observational data. For the synthetic spectra, we set up the following PD and PA values: (30\%; 80°), (10\%; -50°), (25\%; 10°). Table \ref{test} summarizes the tests carried out for each simulated spectrum: the left column lists the models used to synthesize the spectra, while the right column shows the models adopted during the fitting procedure to reproduce the synthetic spectra.

In particular, two different models are tested for this validation process on simulated data: \texttt{Tbabs(vnei+polconst*powerlaw)} and \texttt{Tbabs(polconst*vnei+polconst*powerlaw)} with PD kept fixed to 0 for the thermal component. The red "X" in Table \ref{test} marks the cases where the fitting model is unable to reliably retrieve the input PD and PA values, highlighting key limitations in these configurations. For the tests 2 and 5, the model does not exhibit consistency in reproducing the input values, such as the $\Gamma$ or the PA value. This discrepancy indicates that the model with no thermal component fails to accurately capture the polarization properties. Test 3 should, in principle, retrieve consistent results given the instrinsic zero polarization of the thermal component. However, when fitting a spectra synthesized with the \texttt{polconst} coupled to the thermal component with a model not including the \texttt{polconst} strong discrepancy are again found, despite the PD being set equal to 0. Same occurs in the opposite scenario, Test 6, where the role of the models used to synthesize and to fit the spectra are inverted. These results highlight a likely technical issue either due to the definition of the \texttt{polconst} component in XSPEC or to the IXPE instrumental characteristics.  

The only cases that robustly retrieve the original values occur when the fitting function matches the modeling function, meaning test 1 and test 4. Therefore, we applied the two models \texttt{Tbabs(vnei+polconst*powerlaw)} and \texttt{Tbabs(polconst*vnei+polconst*powerlaw)} to the actual IXPE observations and we noticed that the model \texttt{Tbabs(vnei+polconst*powerlaw)} systematically returns the same PA, no matter the region considered. This suggests a limitation in the model's sensitivity to spatial changes in the source's magnetic field. This spectral model was then discarded. Finally, the second model, \texttt{Tbabs(polconst*vnei+polconst*powerlaw)}, in which the PD coupled to the \texttt{vnei} component is fixed to 0, is chosen for the analysis of the IXPE observations, being able to capture variations in PA across different parts of the SNR, with PA values consistent with the orientations observed in \citet{Vink2022} and with the IXPEOBSSIM unweighted analysis (see Sect. \ref{sec:ixpeobssim}). Although \texttt{polconst*vnei} with \texttt{PD=0} should be equivalent to \texttt{vnei}, this is not the case in practice, likely due to a technical issue in XSPEC. The \texttt{polconst} model is multiplicative and must be applied to all spectral components in the polarimetric fit. When both the thermal and non-thermal components are convoluted with \texttt{polconst}, the fit is correct. Conversely, omitting \texttt{polconst} from the thermal component leads to an incorrect polarization estimate, compromising the analysis of the polarimetric parameters.


\begin{table}
\centering
\tabcolsep=10pt
\begin{tabular}{c|c|c|c}
\hline
\hline
\textbf{Test} & \textbf{Model used to synthesize the spectra synthesis } & \textbf{Model used to fit the synthetic spectra} &  \\ \hline \hline
1 & Tbabs(vnei+polconst*powerlaw) & Tbabs(vnei+polconst*powerlaw) & \textcolor{green}{\checkmark} \\ \hline
2 & Tbabs(vnei+polconst*powerlaw) & Tbabs(polconst*powerlaw) & \textcolor{red}{\texttimes} \\ \hline
3 & Tbabs(vnei+polconst*powerlaw) & Tbabs(polconst*vnei+polconst*powerlaw) & \textcolor{red}{\texttimes} \\ \hline
4 & Tbabs(polconst*vnei+polconst*powerlaw) & Tbabs(polconst*vnei+polconst*powerlaw) & \textcolor{green}{\checkmark} \\ \hline
5 & Tbabs(polconst*vnei+polconst*powerlaw) & Tbabs(polconst*powerlaw) & \textcolor{red}{\texttimes} \\ \hline
6 & Tbabs(polconst*vnei+polconst*powerlaw) & Tbabs(vnei+polconst*powerlaw) & \textcolor{red}{\texttimes} \\
\hline
\hline
\hline
\end{tabular}
\caption{Comparison between the models used to generate synthetic spectra and those employed for fitting them. Green checkmarks indicate that the approach successfully recovers the parameter values used for the spectral synthesis. Red Xs denote models providing parameter values having significant discrepancy with the input parameters.} 
\label{test}
\end{table}

\subsection{IXPEOBSSIM Unweighted analysis}
\label{sec:ixpeobssim}
As a further sanity check, we also compared the spectropolarimetric results with the standard analysis using the open-source software package \texttt{IXPEOBSSIM} \citep{Baldini2022}, utilizing version 12 of the IXPE response functions and adopting unweighted analysis approach.
The data from IXPE’s three DUs are combined by processing the individual level 2 event files, corrected for instrumental background, through the \texttt{xpbin} tool in \texttt{IXPEOBSSIM}. Using the \texttt{PMAPCUBE} algorithm within \texttt{xpbin}, we bin the polarization maps with a pixel size of 42'', for consistency with the spectral extraction regions, focusing on the 3–6 keV energy band. Afterward, using \texttt{xpselect}, we select each region of interest to derive PD and PA through the \texttt{PCUBE} algorithm within \texttt{xpbin}, subtracting the background and subsequently correcting for the corresponding synchrotron fraction (see Section B in the Appendix).

\section{RESULTS}
\subsection{Polarization measurements}
The PD and PA values obtained from the analysis are reported in Table \ref{tab:PD_PA} with the corresponding uncertainty (at 1$\sigma$ confidence level) and the significancy of the detection, estimated for one degree of freedom, as it is standard practice for IXPE analysis \citep{Ferrazzoli2023,Zhou2023,Ferrazzoli2024,Prokhorov2024}. We also show in Fig. \ref{mappe} a map of PD and PA for all the regions in which the detection of X-ray polarization is significant at least at the $2\sigma$ (yellow vectors) or $3\sigma$ (red vectors) levels.
Five regions (specifically regions 8, 9, 10, 21, and 22) are excluded a-priori due to overwhelming thermal contamination observed in the Chandra spectra. This exclusion leaves 18 regions for our analysis. For the most significant region (region 1), the post-trial significance is \( (99.987\%)^{18} = 99.77\% \), namely 3.05$\sigma$. PD varies significantly in different areas of the shell, ranging from $\sim 10\%$ to $\sim 26\%$, with the highest and most-significant values in the northeast and southeast. Notably, these values are generally higher than those reported by \citet{Vink2022}, likely due to the different estimation methods of the thermal contribution. In our case, the thermal contribution was estimated through the analysis of Chandra spectra, whereas in \citet{Vink2022}, the PD was corrected for thermal contamination using an estimate of the ratio between thermal and non-thermal fluxes. 
In Fig. \ref{mappe}, it can also be observed that the orientation of the polarization vectors is mostly parallel to the shock front, indicating a radially-oriented magnetic field, in agreement with the radio observations \citep{Rosenberg1970,Braun1987} and the \texttt{IXPEOBSSIM} analysis by \citet{Vink2022}. This result suggests, as one may expect, that the contribution from thermal emission has an impact only on the measured PD and not on the PA. To further verify the orientation, a spectropolarimetric fit is performed after extracting the spectra for each region from the aligned data using the \texttt{IXPEOBSSIM} tool \texttt{xpstokesalign}, assuming tangential polarization. If the PA is close to 0, the polarization can be considered parallel to the shock. The obtained PA values are consistent with those found in the spectropolarimetric analysis. 
\subsection{Comparison with \texttt{IXPEOBSSIM} analysis}

\begin{figure}[h!]
    \centering
    \includegraphics[width=1.0\linewidth]{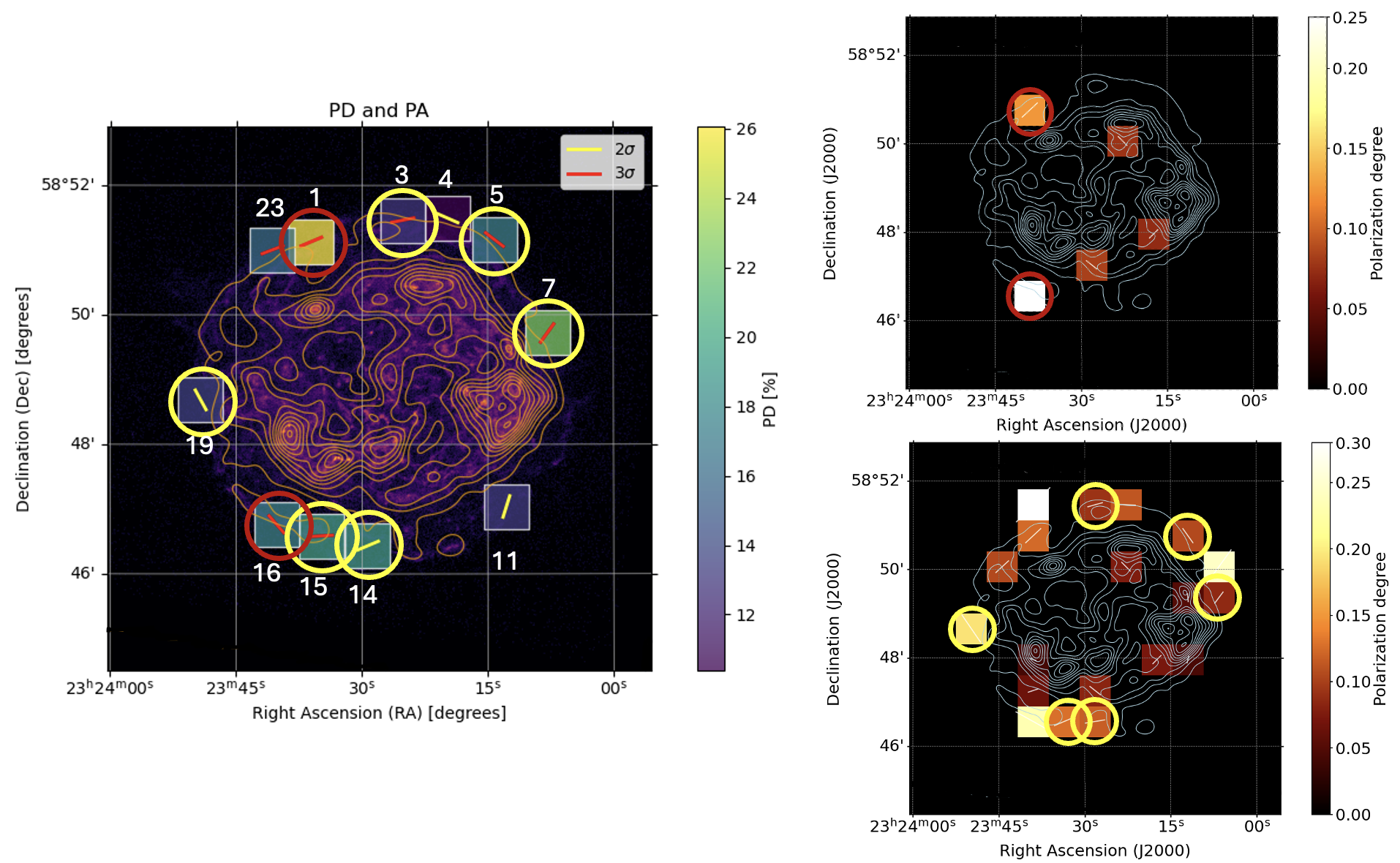}
    \caption{\textit{Left panel.} Map of PD and PA values produced by the spectropolarimetric analysis in the 3–6 keV energy band, overlaid with vectors indicating X-ray polarization detections. Red and yellow vectors correspond to regions with significant detections at the 3- and 2-$\sigma$ confidence level, respectively. \textit{Right panels.} Polarization maps obtained through the \texttt{IXPEOBSSIM} analysis, with the top and bottom images highlighting pixels where detections exceed the 3$\sigma$ and 2$\sigma$ confidence levels, respectively. Regions with detections exceeding  the 3$\sigma$ and 2$\sigma$ significance levels in the \texttt{IXPEOBSSIM} that are detected by both the analysis methods are marked with red and yellow circles, respectively. Notice that the regions shown on the left and right panels do not completely overlap due to the automatic selection of them of \texttt{IXPEOBSSIM}. This does not affect the results because of the large IXPE PSF.}
    \label{mappe}
\end{figure}

In Fig. \ref{mappe}, we present a comparison between the spectral analysis map (left) and the polarization maps obtained through the \texttt{IXPEOBSSIM} analysis (right). The top-right image highlights pixels where detections exceed the 3$\sigma$ significance level, while the bottom-right image shows pixels with detections above the 2$\sigma$ level.
Many regions in the spectral map on the left find counterparts in the polarization maps on the right. Among them, regions with detections above 3$\sigma$, according to the \texttt{IXPEOBSSIM} analysis, are marked with red circles, while those above 2$\sigma$ are indicated with yellow circles. There are some discrepancies in the positioning of regions between the two maps, which can be entirely attributed to the pixel centering. An important finding is that our \texttt{IXPEOBSSIM} maps show a different number of significant pixels compared to those reported in Figure 3 in \citet{Vink2022}. This discrepancy can be attributed to the updates of the \texttt{IXPEOBSSIM}. 
It is worth noticing that the 2 dof analysis adopted by \citet{Vink2022} has led to a polarization degree for a pixel size of $42"$ in the outer regions of $\sim 19 \%$, which is comparable with our results from the \texttt{IXPEOBSSIM} analysis, after synchrotron fraction correction.
In Table \ref{tab:PD_PA} we report the polarization values for each region of interest derived through the \texttt{PCUBE} algorithm within \texttt{xpbin}, correcting for the background, computing PD and PA from the relevant Stokes parameters, and then correcting again for the corresponding synchrotron fraction (see Appendix B), alongside the spectropolarimetric results. The results obtained from both analyses show a strong overall similarity. However, some differences emerge in specific cases. For instance, in regions 7, 14, and 23, the \texttt{IXPEOBSSIM} analysis, despite accounting for the synchrotron fraction correction, results in a less significant polarization detection with respect to the spectropolarimetric analysis. The values of PA are remarkably in agreement between the two types of analysis for all the regions selected, confirming that the PA is marginally affected by thermal contamination.

\begin{table} [h!]
    \centering
   \tabcolsep=10pt
    \begin{tabular}{c|c|c|c|c|c|c|c}
    \hline
     \hline
    \multicolumn{1}{c|}{} & \multicolumn{3}{|c|} {spectropolarimetric analysis} & \multicolumn{4}{|c} {\texttt{IXPEOBSSIM} analysis} \\
    \hline
   
        Region & PD ($\%$) & PA (°) & $\sigma$ & PD ($\%$) & PD$_{corr}$ ($\%$) & PA (°) & $\sigma$ \\
        \hline
         1 & 26.06$_{-6.76}^{+6.80}$ & -67.21$_{-7.52}^{+7.47}$ & 3.83 & 20.56$_{-3.82}^{+3.82}$ & 22.98$_{-4.27}^{+4.27}$ & -71.54$_{-7.12}^{+7.12}$ & 4.02 \\
         3 & 12.91$_{-4.18}^{+4.19}$ & -78.64$_{-9.40}^{+9.45}$ & 3.08 & 10.64$_{-3.82}^{+3.82}$ &  12.39$_{-4.45}^{+4.45}$ & -75.95$_{-10.27}^{+10.27}$ & 2.79 \\
         4 & \(<\)22.31 & - & 2.25 & \(<\)22.45 & \(<\)25.40 & - & 2.36 \\
         5 & 17.32$_{-4.93}^{+4.95}$ & 52.54$_{-8.19}^{+8.25}$ & 3.50 & 17.16$_{-4.52}^{+4.52}$ &  19.20$_{-5.06}^{+5.06}$ & 47.65$_{-7.54}^{+7.54}$ & 3.80 \\
         7 & 22.47$_{-6.78}^{+6.81}$ & -35.68$_{-8.67}^{+8.64}$ & 3.30 & 10.14$_{-3.95}^{+3.95}$ & 11.56$_{-4.50}^{+4.50}$ & -36.19$_{-11.16}^{+11.16}$ & 2.57 \\
         11 & \(<\)28.35 & - & 2.24 & \(<\)25.23 & \(<\)29.19 & - & 2.10\\
         14 & 19.21$_{-7.21}^{+7.25}$ & -67.14$_{-11.03}^{+11.09}$ & 2.66 & \(<\)28.05 & \(<\)30.82 & - & 2.28 \\
         15 & 19.21$_{-5.92}^{+5.94}$ & -86.60$_{-8.93}^{+8.93}$ & 3.24 & 18.57$_{-5.62}^{+5.62}$ & 20.38$_{-6.17}^{+6.17}$ & -81.19$_{-8.68}^{+8.68}$ & 3.30 \\
         16 & 17.50$_{-5.44}^{+5.45}$ & 41.00$_{-9.07}^{+9.14}$ & 3.21 & 15.59$_{-5.75}^{+5.74}$ & 17.03$_{-6.28}^{+6.28}$ & 40.72$_{-10.57}^{+10.57}$ & 2.71 \\
         19 & \(<\)28.93 & - & 2.24 & \(<\)26.07 & \(<\)30.79 & - & 2.42 \\
         23 & 16.09$_{-5.28}^{+5.30}$ & -69.03$_{-9.49}^{+9.65}$ & 3.04 & \(<\)24.49 & \(<\)26.87 & - & 2.27 \\
         \hline
         \hline
    \end{tabular}
    \caption{Results from the spectropolarimetric analysis and the \texttt{IXPEOBSSIM} analysis, the latter after correction for the corresponding synchrotron fraction (PD$_{corr}$), of some regions reported in Fig. \ref{mappe}. The uncertainties reported are at the 1 $\sigma$ confidence level. The \texttt{IXPEOBSSIM} analysis is determined assuming 1 degree of freedom. For regions where the significance of the measurement is below 2.57$\sigma$, the upper limit for PD is reported at the 99\% confidence level, while the PA is non-constrained.}
    \label{tab:PD_PA}
\end{table}

\subsection{Estimation of the $\eta$ factor}\label{sec:eta}
\begin{figure}
    \centering
    \includegraphics[width=12.5cm]{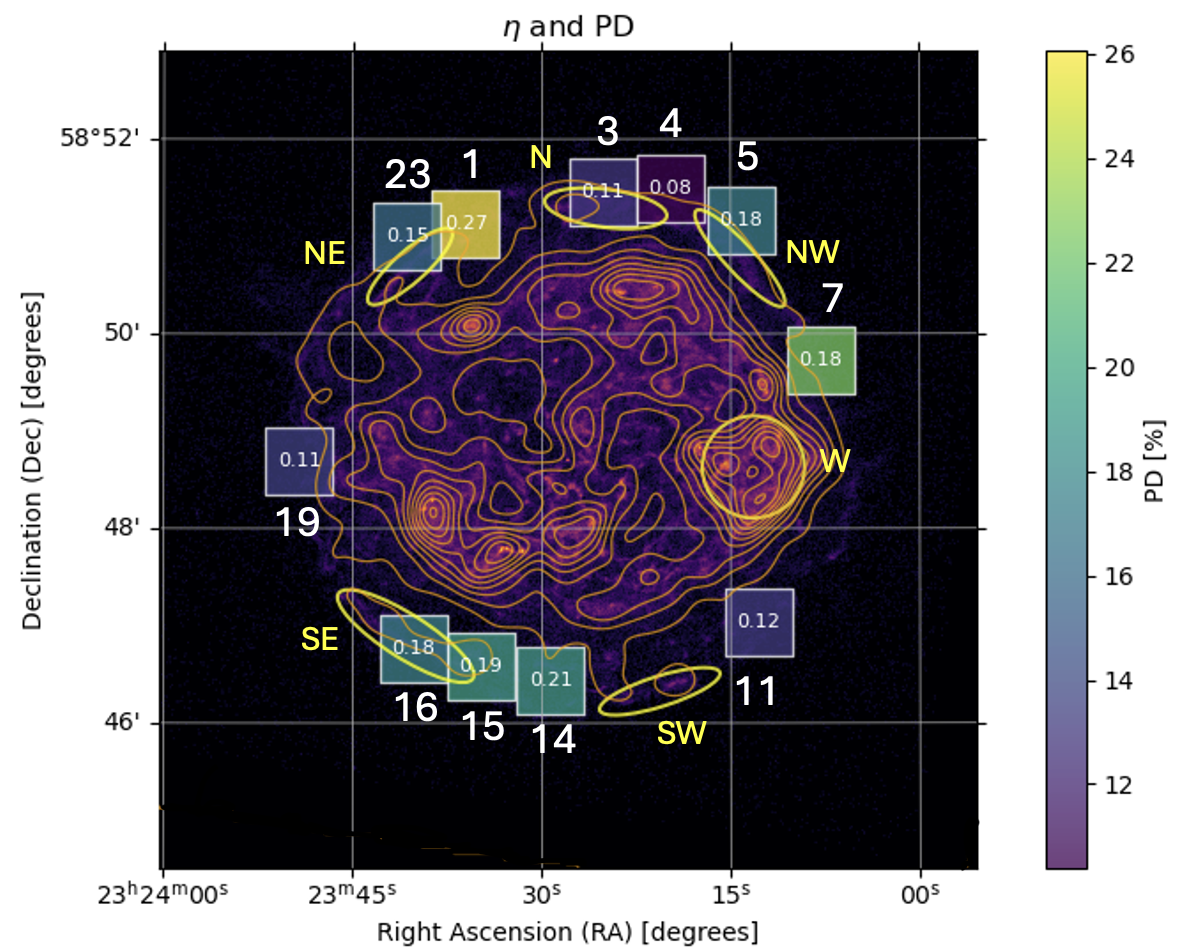}
    \caption{Map of the degree of magnetic turbulence $\eta$ and of the PD from the spectropolarimetric analysis in the 3-6 keV energy band for all the regions where the detection of X-ray polarization is significant at least at the 2$\sigma$ confidence level. The corresponding $\eta$ value is displayed in the box for each region. Yellow ellipses and corresponding labels mark the regions analyzed by \citet{Greco2023} (see Sect. \ref{sec:jitter}).}
    \label{PD_eta}
\end{figure}

The degree of polarization is strictly linked to the presence of ordered or disordered magnetic field. It is well known that DSA requires highly turbulent ambient fields, so that it is interesting to make an attempt to link the PD coming from the IXPE analysis to the degree of disorder of the magnetic field. In order to do so, we follow the derivation described in \citet{Bandiera2016}, where they found that, under the assumption of isotropic magnetic field fluctuations, the PD is given by \citep{Bandiera2016,Bandiera2024} 
\begin{equation}
PD =  \frac{s+1}{s+\frac{7}{3}} \frac{(5+s)}{8} \frac{3 \bar{B}^2}{2 \delta B^2} \frac{{}_1F_1 \left( \frac{(3-s)}{4}, 3, -\frac{3\bar{B}^2}{2 \delta B^2} \right)}{{}_1F_1 \left( \frac{-(1+s)}{4}, 1, -\frac{3 \bar{B}^2}{2 \delta B^2} \right)},
\label{eq}
\end{equation}
where $s = 2\Gamma - 1 $ is the spectral index, $\Gamma $ is the photon index obtained from Chandra data analysis, and ${}_1F_1$ is the Kummer confluent hypergeometric function.  This is indeed, a very simplified assumption since it does not take into account any turbulent cascade. Notice also that the assumption of a downstream isotropic fluctuating field and the observations of a mainly radial large scale field can be difficult to reconcile; thus, the following estimation of the ratio between the disordered over the average magnetic field should be taken with caution, since the intrinsic anisotropy of the field downstream, due to the shock compression, can actually change those values \citep{Ferrazzoli2023,Zhou2023}. \\
In the case of highly turbulent fields, namely $\delta B>>B$, the asymptotic expression for equation (\ref{eq}) is simplified to
\begin{equation}
  PD = \frac{s+1}{s+\frac{7}{3}} \frac{5+s}{8} \frac{3 B^2}{2 \delta B^2}.
\end{equation}
 Once the $PD$ is determined by the IXPE data analysis, we can calculate the degree of magnetic turbulence, $\eta$, from the full equation (\ref{eq}), which quantifies the ratio between the mean magnetic field and its fluctuating part, namely 
\begin{equation}
    \eta = \left( \frac{B}{\delta B} \right)^2.
\end{equation}
The values of $\eta$ are provided in Table \ref{eta} for each analyzed region in the outer shell of Cas A. Fig. \ref{PD_eta} shows the $\eta$ factor and PD for regions with detection $>2\sigma$ and $>3\sigma$. All the regions show values of $\eta < 1$, indicating that, despite some deviations are found across different regions, the downstream plasma is always characterized by a fluctuating magnetic field $\delta B$ much stronger than the mean one $B$. Our results are consistent with those reported by \citet{Ferrazzoli2023} for Tycho's SNR, as they also find $\delta B/B > 1$. Specifically, they report $\eta = 0.09 \pm 0.02$ at the rim and $\eta = 0.21 \pm 0.08$ in the western region.
We stress that $\eta$ here should not be mistaken for the Bohm factor, as it is not a metric of the efficiency of the acceleration but only on the level of disordered of magnetic fields in the downstream plasma. In conclusion, in Cas A 
the anisotropy induced by the mean field is effectively reduced by the dominant turbulent fluctuations \citep{Wentzel1974}. 
\begin{table}
    \centering
    \tabcolsep=13pt
    \begin{tabular}{c|c}
       \hline
       \hline
       Regions  & $\eta$ \\
       \hline
        1 & 0.27$_{-0.09}^{+0.11}$ \\
        3 & 0.11$_{-0.04}^{+0.04}$ \\
        4 & 0.08$_{-0.04}^{+0.05}$ \\
        5 & 0.18$_{-0.06}^{+0.07}$ \\
        7 & $<0.18$ \\
        11 & 0.12$_{-0.06}^{+0.07}$ \\
        14 & 0.21$_{-0.09}^{+0.11}$ \\
        15 & 0.19$_{-0.07}^{+0.08}$ \\
        16 & 0.18$_{-0.07}^{+0.08}$ \\
        19 & 0.11$_{-0.05}^{+0.06}$ \\
        23 & 0.15$_{-0.06}^{+0.07}$ \\
        \hline
        \hline
    \end{tabular}
    \caption{Degree of magnetic turbulence values ($\eta$) for the regions shown in Fig. \ref{PD_eta}, with their respective uncertainties.}
    \label{eta}
\end{table}

\section{Discussion}

This study presents a detailed spectropolarimetric analysis of the SNR Cas A using IXPE data, combined with Chandra observations in order to take into account the thermal contamination in the emission.
Accurate estimation of the thermal contribution allows to better isolate the polarized emission, as the thermal component dilutes the intrinsic polarization of the non-thermal emission.
PD values, determined by properly modeling such contamination, range locally between 10\% and 26\%, remarkably higher than the average ones reported by \citealt{Vink2022}, but still much lower than the intrinsic limit of $\sim 75\%$, indicating a highly complex magnetic field morphology even when considering small-scale areas.

\subsection{On the polarization degree and angle} 
PD reaches its peak in the southeastern and northeastern regions. The PD inferred locally through the spectral and the IXPEOBSSIM unweighted analysis is higher than, but generally compatible with, the one previously reported by \citet{Vink2022}, with values that locally are even similar to those reported by \citet{Zhou2023} for SN1006. This is mainly ascribable to three factors: i) the updates on the IXPEOBSSIM software and the improved data reduction and background subtraction techniques; ii) the inclusion of an accurate thermal component in the spectral analysis; iii) the size of the regions analyzed. In fact, while \citet{Vink2022} did report an average polarization fraction of $\sim 5\%$, that was achieved by considering significantly larger regions than the boxes here chosen: in the map from \citet{Vink2022}, the PD can be as high as $20\%$ in a couple of pixels. Averaging across the whole shell, as done by \citet{Vink2022} may include regions with lower PD. In any case, the number of significant regions in our analysis is slightly higher and, most importantly, the level of PD in each pixel is systematically higher thanks to the inclusion of a proper thermal component in the spectropolarimetric analysis.

Regions 3, 4 and 5 located in the northwestern part of the SNR are characterized by a relatively lower PD with respect to the northeastern and southeastern areas. Interestingly, the northwestern part of Cas A is where interaction between the forward shock and a shell of dense CSM occurred $\sim 200$ yr after the SN event, as invoked to explain the inward movement of the reverse shock in the West \citep{Vink2022b,Orlando2022}. The interaction between the forward shock and dense shell generates a secondary reverse shock which moves inwards in the observer rest of frame. The motion of this secondary reverse shock has two effects on the magnetic field: i) it is stretched and elongated as the forward shock expands in the ambient medium resulting in a particularly enhanced radial magnetic field orientation, particularly visible in region 5; ii) the shell is expected to present an overall clumpiness that is enhanced after the interaction with the shock, leading to the formation of instabilities and turbulence which decrease the level of polarization. These two effects, mixed, lead to region 5 as the one with the best-constrained PA and, at the same time, to an overall lower PD with respect to the northeastern and southeastern arcs, where no interaction has occurred. 

The polarization maps presented in Fig. \ref{mappe} suggest that in many regions, the polarization vectors are aligned with the shock front, implying a predominantly radial magnetic field. This pattern is in agreement with the previous IXPE study by  \citet{Vink2022}. This is at odds with what one would expect when considering the shock-compression effects on the magnetic field, leading to a an orientation mainly tangential. However, \citet{Bykov2024}, by means of magnetohydrodynamic (MHD) simulations, where the effect of the electric current induced by accelerated protons was also included, interpreted the emergence of a radial magnetic field (i.e., a tangential polarization of the emission) as an effect of anisotropy of magnetic turbulence far downstream of the shock: just behind the shock the transverse magnetic field dominates due to the shock compression. Thus, a distribution of multi-TeV electrons confined within that region would give rise to a X-ray longitudinally polarized emission. On the other hand, the radial component of the magnetic field, amplified via a small-scale turbulent dynamo mechanism, starts being dominant far downstream, producing the transverse-polarized synchrotron emission seen in X-ray observations. 

The only regions which show non radial magnetic fields are regions 7 and 11, in the West. This is the area of Cas A where the counter-jet lies and where the reverse shock is producing bright non-thermal emission. Though a more detailed modeling would be needed to properly take into account of the interplay between different factors, it is not surprising that such a complex environment leads to the formation of structure radically different from the other areas of the SNR.

\citet{Vink2022} reported an average PD for the forward shock region of $\sim 5\%$, comparable to, and in some regions even lower than, the average radio PD. This result is surprising, since X-ray synchrotron emission originates from thinner regions downstream of the shock, minimizing depolarization caused by magnetic field variations, while radio emission arises from larger volumes with significant depolarization along the line of sight. Additionally, the steep X-ray spectrum implies a higher intrinsic PD than in the radio band. Here, we find several regions with X-ray PD higher not only to the previous IXPE analysis but also to the radio band \citep{Rosenberg1970,Braun1987}. The inclusion of an accurate modeling of the thermal emission is therefore crucial to properly reconcile the two bands. The spectropolarimetric and \texttt{IXPEOBSSIM} methods are generally consistent, except for the significance of the measurement with the unweighted analysis of \texttt{IXPEOBSSIM}, which is lower than that obtained with the spectropolarimetric analysis. However, slight differences are observed in some areas, such as Regions 7, 14, and 23, where the \texttt{IXPEOBSSIM} approach yields marginally lower PD values and reduced significance compared to the spectropolarimetric results. 

\subsection{Polarization and jitter radiation}
\label{sec:jitter}
In Fig. \ref{PD_eta}, we show a map of the degree of magnetic turbulence $\eta$, i.e. the inverse of the ratio between the mean magnetic field and its fluctuating part, with values ranging between 0.08 and 0.27. On one hand this highlights a variation in the level of magnetic turbulence but, on the other hand, $\eta$ is anyway below 1, indicating a fluctuating magnetic field dominant over the mean one ($\delta B >> B$).

It is not obvious to discriminate between the turbulent magnetic field $\delta B$ and the mean magnetic field $B$ when the fluctuations can be even greater than the actual absolute value of $B$. A possibility is that during its motion along the magnetic field lines, the electrons trajectories are distorted by the turbulent magnetic field before they complete a gyro-orbit. This scenario, in which turbulence on the small scales affects the emission process of the synchrotron radiation, was recently investigated by \citet{Greco2023} who adopted the jitter radiation model developed by \citet{Toptygin1987} and \citet{Kelner2013} and applied it to Cas A. Jitter radiation is the extension of synchrotron radiation when the magnetic field in which the electron is embedded is not uniform (see \citealt{Greco2023} and reference therein for additional detail)\footnote{According to \citet{Kelner2013}, for SNRs it is needed that the minimum scale of the turbulence is $\lambda << 170 $ km, for jitter radiation to be at work. This is a scale much larger than the dissipative ones: for example, Landau damping acts on scales smaller than 1 km}. Being originated by the turbulent magnetic field, it is intrinsically unpolarized and its spectrum of emission is characterized by a break: photons below the break obey to the standard synchrotron radiation, and are expected to be polarized, whereas photons above the break are originated by the turbulent magnetic field. Therefore, we expect that regions with higher break energy have an overall higher polarization, since the amount of photons emitted in the synchrotron regime is larger with respect to a case with lower break energy. \citet{Greco2023} analysed 6 regions across the shell: Fig. \ref{PD_eta} shows the cross-match between them and the boxes used in this analysis. Remarkably, all the regions characterized by energy break higher than 5 keV have at least one match with one of the 3$\sigma$ detections. Moreover, the region SW, which was completely dominated by the jitter regime, is the only one showing no significant detection. The only exception is raised by the W region, which does not show high polarization signal. However, W region of \citet{Greco2023} has no real counterpart in the spectropolarimetric analysis, since it is associated with the synchrotron emission from the reverse shock. The reverse shock may be much more irregular than the forward shock, which could cause the PAs to become scrambled, leading to a lower PD. 
Overall, the jitter radiation scenario and the spectropolarimetric results presented in this paper align fairly well, reflecting the crucial importance of magnetic turbulence in the emission process of synchrotron radiation.


\section{Conclusions}
The main findings of the combined Chandra and IXPE data analysis of the outer shell of Cas A can be summarized as follows:
\begin{itemize}
    \item \citet{Vink2022} only reported average PDs, giving a maximum PD of 5\%. Here we show that the PD can locally go as high as 25\% in regions close to the forward shock, while the PA consistently aligns with a radial magnetic field. Regions 7 and 11, close to the counter-jet regions, are the only ones that show tangential magnetic field, possibly because of the local complex dynamics and configuration related to the counter-jet and future dedicated modeling could help in understanding these features.

    \item The northeastern and southeastern regions of the shell exhibit the highest polarization values. The regions with the highest significance are Region 1, in the bright nonthermal arc in the northeast, (3.83$\sigma$) and Region 5 in the northwest, close to the region of interaction between a dense CSM shell and the forward shock (3.50$\sigma$). In such a region indeed the value of PD is low, probably because such an interaction can give rise to turbulent magnetic fields.

    \item While being overall in agreement with the IXPEOBSSIM analysis, the spectropolarimetric approach turned out to be able to capture a higher number of regions with significant X-ray polarization, thanks to the proper modeling of the underlying thermal emission. On the other hand, the results on the PA are perfectly in agreement between two methods, indicating that the presence of the thermal component only affects the polarization degree but not its orientation.

    \item The degree of magnetic turbulence $\eta$, which estimates the ratio between the mean magnetic field and its fluctuations, has provided insights into the level of magnetic turbulence in the outer shell of Cas A. Its values, varying between 0.08 and 0.27, indicate a very high level of turbulence levels across the remnant. Our findings align with those of \citet{Ferrazzoli2023} for Tycho's SNR, as their results also indicate $\eta < 1$. In this work we have used the assumption of isotropic magnetic field fluctuations \citep{Bandiera2016}, it would be probably interesting to estimate $\eta$ by assuming other turbulent field configurations.

\end{itemize}

\section*{Acknowledgements}
 We thank the anonymous referee for the prompt report and its useful suggestions that helped improved the paper.  E.G. thanks M. Miceli, S. Orlando and V. Sapienza for fruitful discussion on the nature and dynamics of the dense shell of CSM in Cas A. This research used data products provided by the IXPE Team (MSFC, SSDC, INAF, and INFN). It was distributed with additional software tools by the High-Energy Astrophysics Science Archive Research Center (HEASARC) at NASA Goddard Space Flight Center (GSFC). R.F. is supported by the Italian Space Agency (Agenzia Spaziale Italiana, ASI) through agreement ASI-INAF-2022-19-HH.0 with the Istituto Nazionale di Astrofisica (INAF), and by MAECI with grant CN24GR08 “GRBAXP: Guangxi-Rome Bilateral Agreement for X-ray Polarimetry in Astrophysics”. 
 S.P. acknowledges Space It Up project funded by the Italian Space Agency, ASI, and the Ministry of University and Research, MUR, under contract n. 2024-5-E.0 - CUP n. I53D24000060005.

\begin{center}
    APPENDIX
\end{center}
\begin{appendices}
\section{2 dof analysis}
\begin{figure} [!h]
   \centerline {\includegraphics[width=18cm]{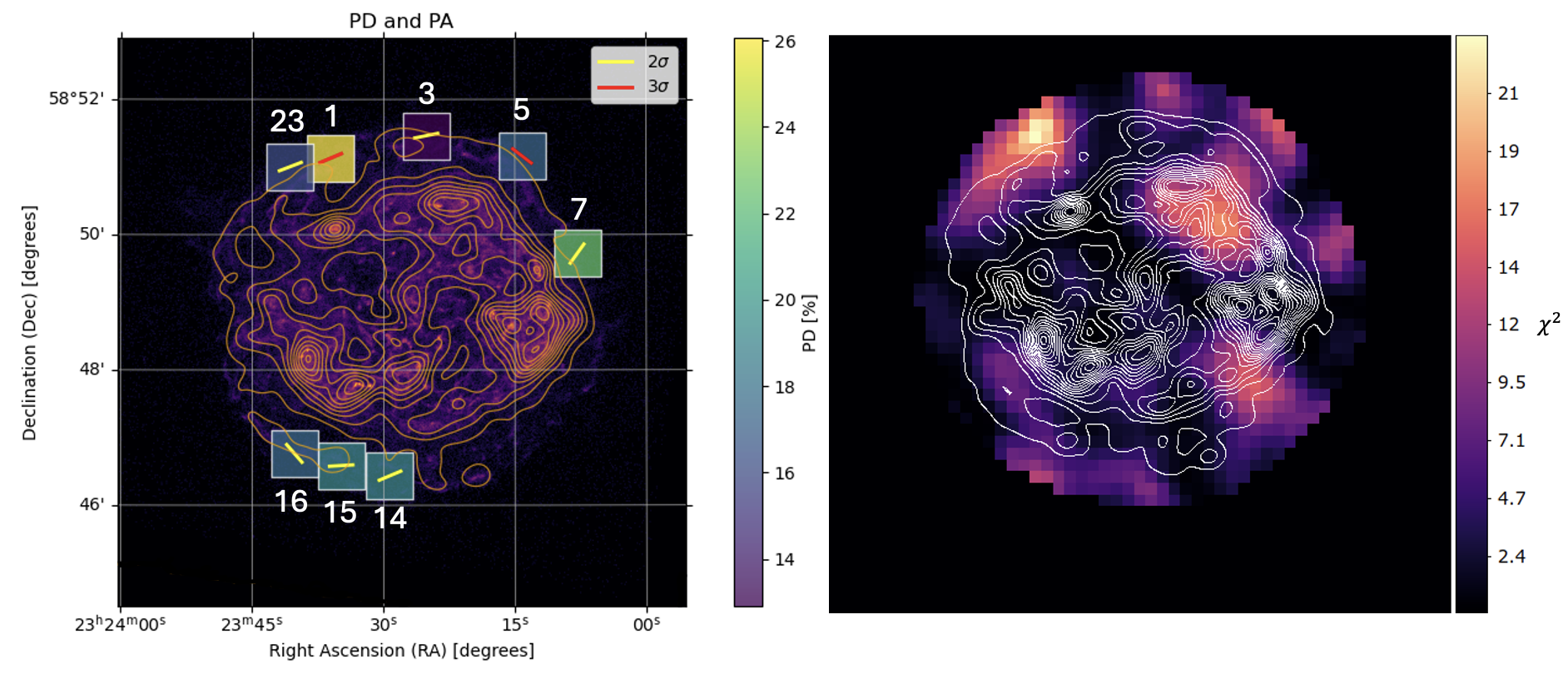}}
    \caption{Left: Map of PD and PA values, produced by spectropolarimetric analysis in the 3-6 keV energy band, for all regions where the detection of X-ray polarization is significant at least at the 2$\sigma$ CL (yellow vectors) or 3$\sigma$ CL (red vectors), with the CL determined considering the droplet-contours.
    Right: Map of $\chi^2$ values  for the polarization signal for the 3–6 keV band, smoothed using the Welch filter, with kernel size of 11 pixels.}
    \label{PD_PA_2}
\end{figure}

To assess whether there is evidence for X-ray polarization we can use the $\chi^2_2$ test (e.g. \citealt{Vink2022}). If the $\chi^2$ value is above a certain value (e.g. 9.2 for 99\% confidence) we can reject the null-hypothesis that there is no polarization. However, for interpretation of the measurements we are interested in the PD versus PA, which are two non-orthogonal quantities derived from the orthogonal quantities Q and U. The confidence regions for PD and PA can be visualized in a two dimensional plot with a characteristic droplet-like shape\citep{Vink2022, Ferrazzoli2023, Zhou2023, Ferrazzoli2024, Prokhorov2024}. If a confidence contour of a certain confidence level (CL) just touches the origin of the polar plot, the data are consistent with the detection of polarization at this specific CL.

\begin{figure}
    \includegraphics[width=0.34\linewidth]{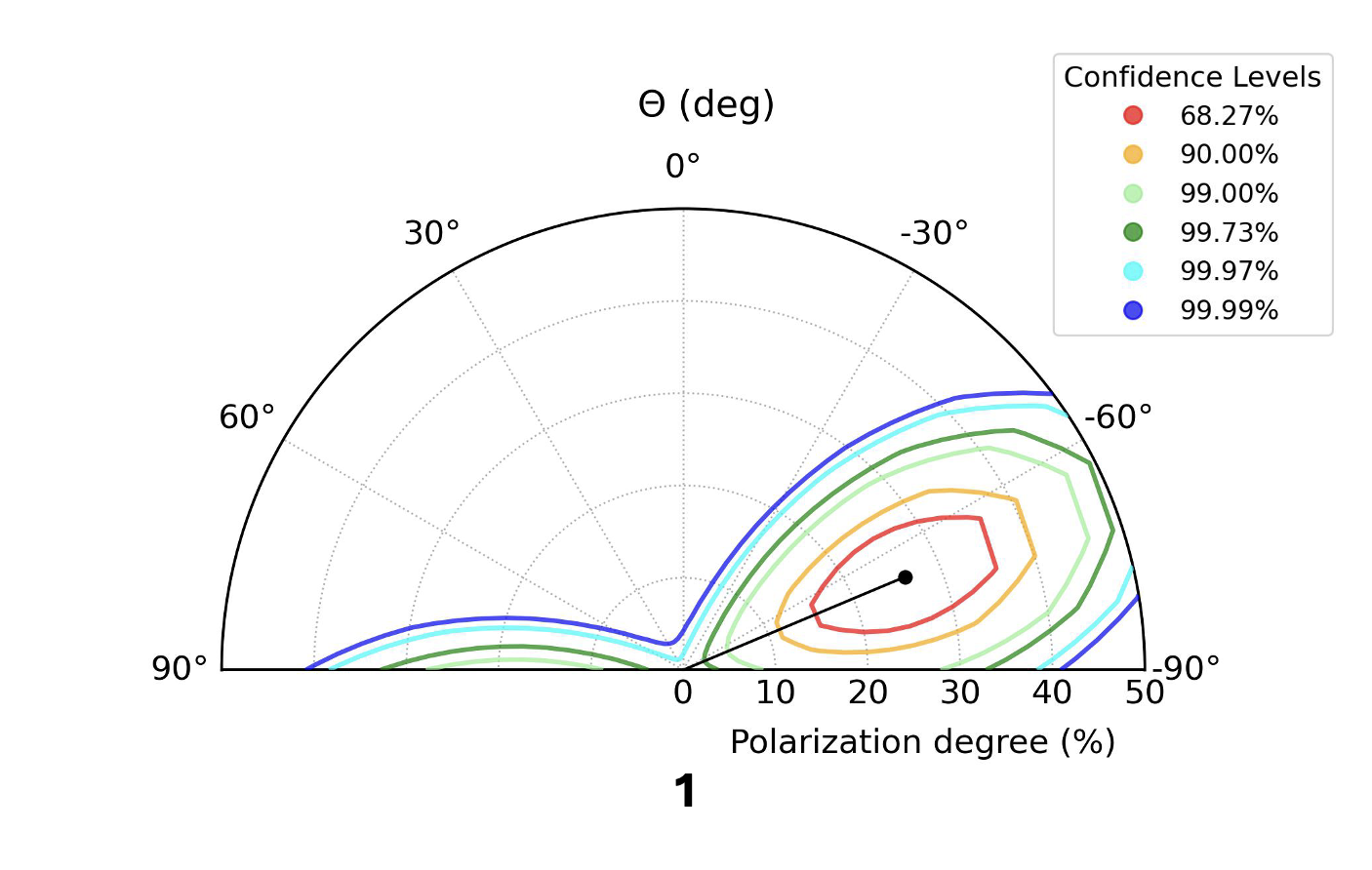}
    \includegraphics[width=0.34\linewidth]{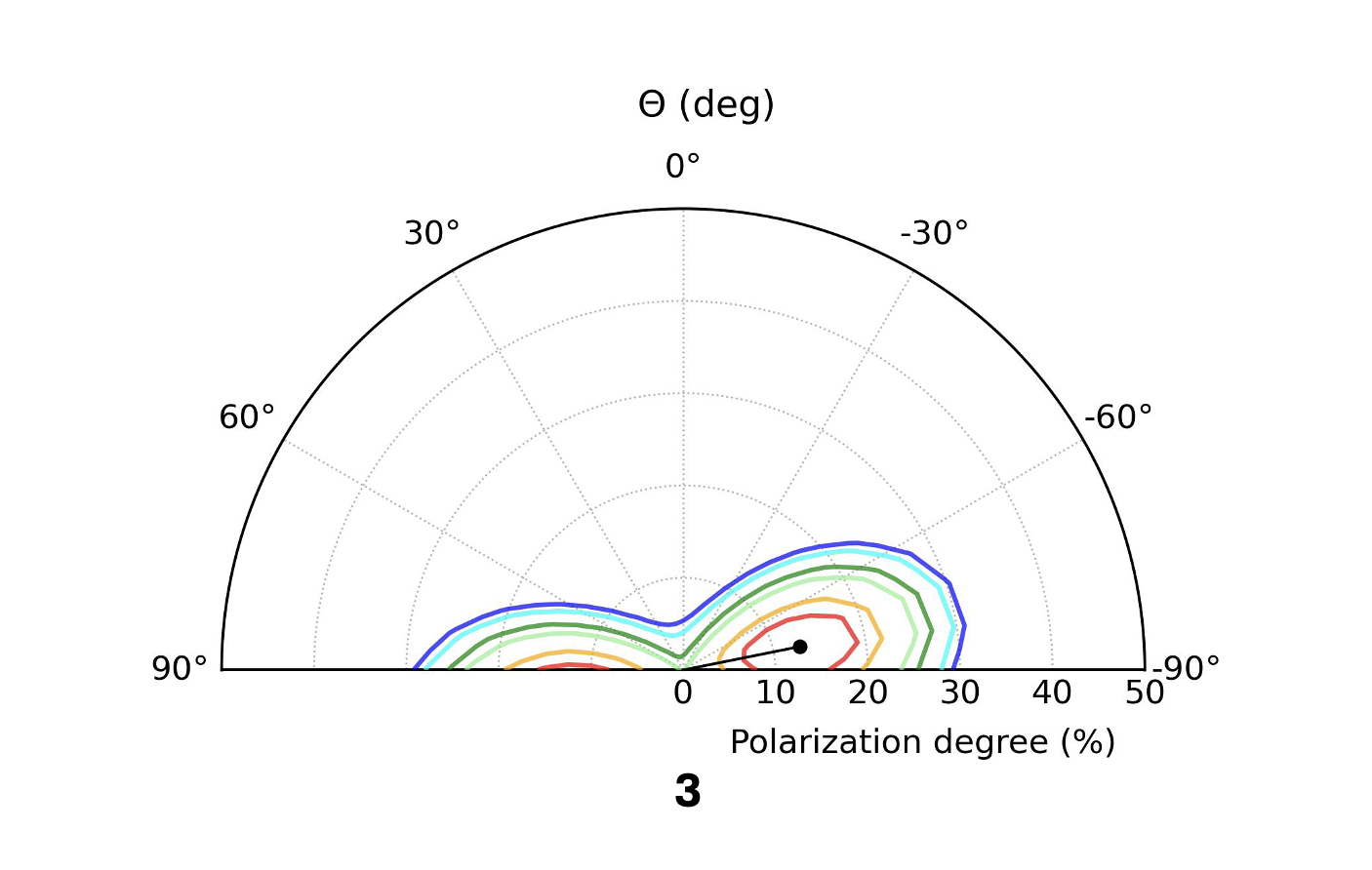}
    \includegraphics[width=0.34\linewidth]{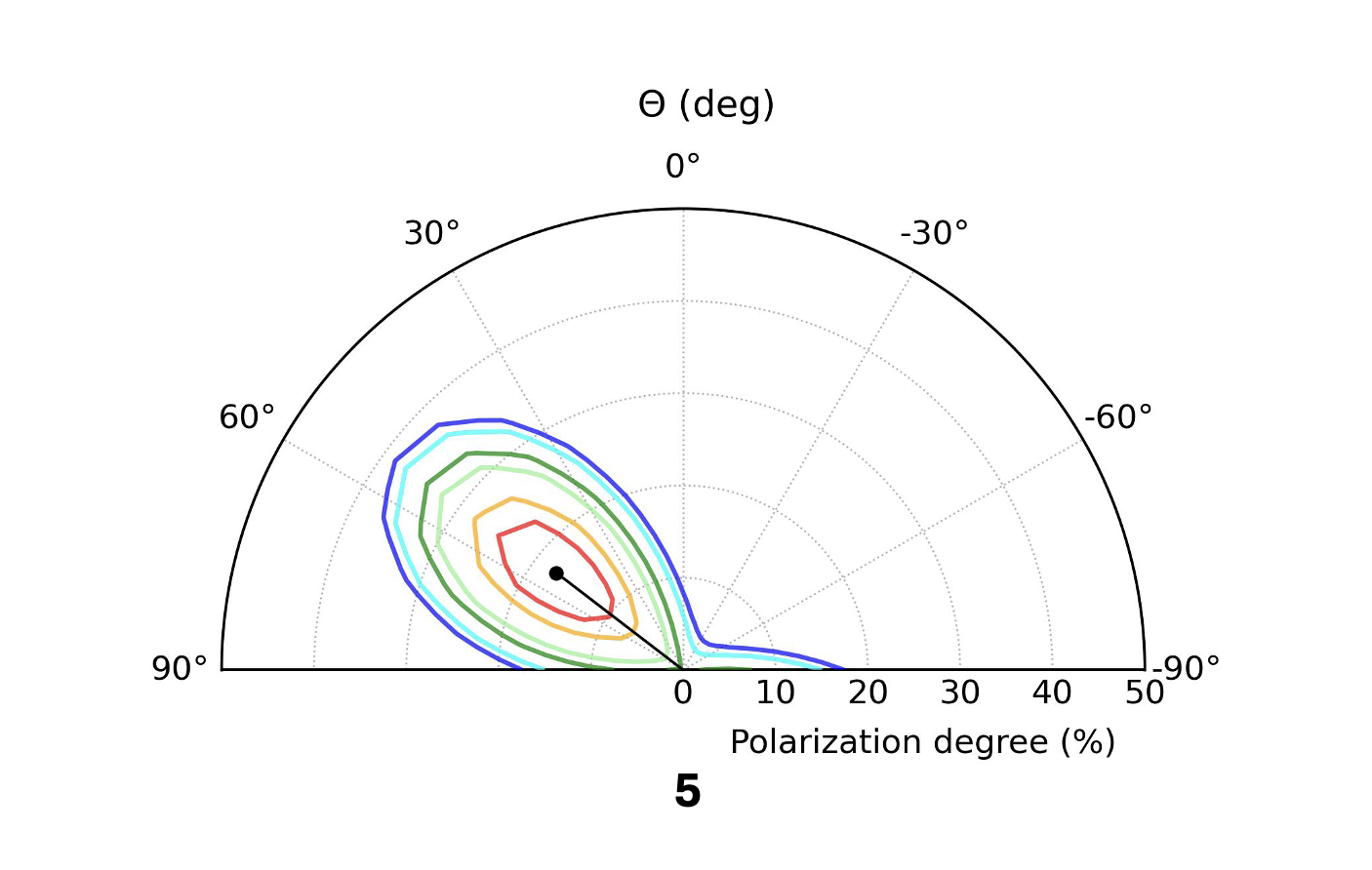}
    \includegraphics[width=0.34\linewidth]{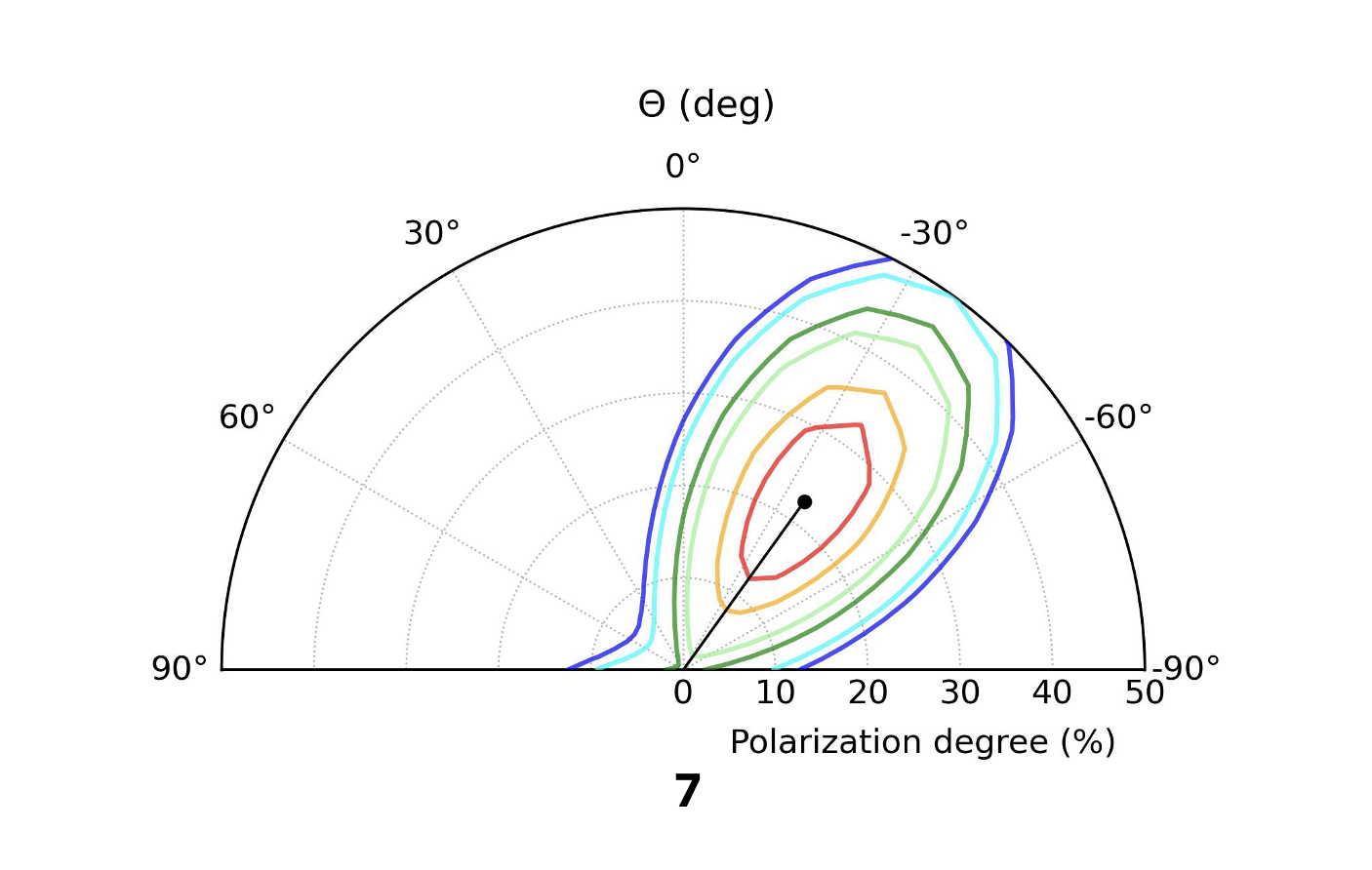}
     \includegraphics[width=0.34\linewidth]{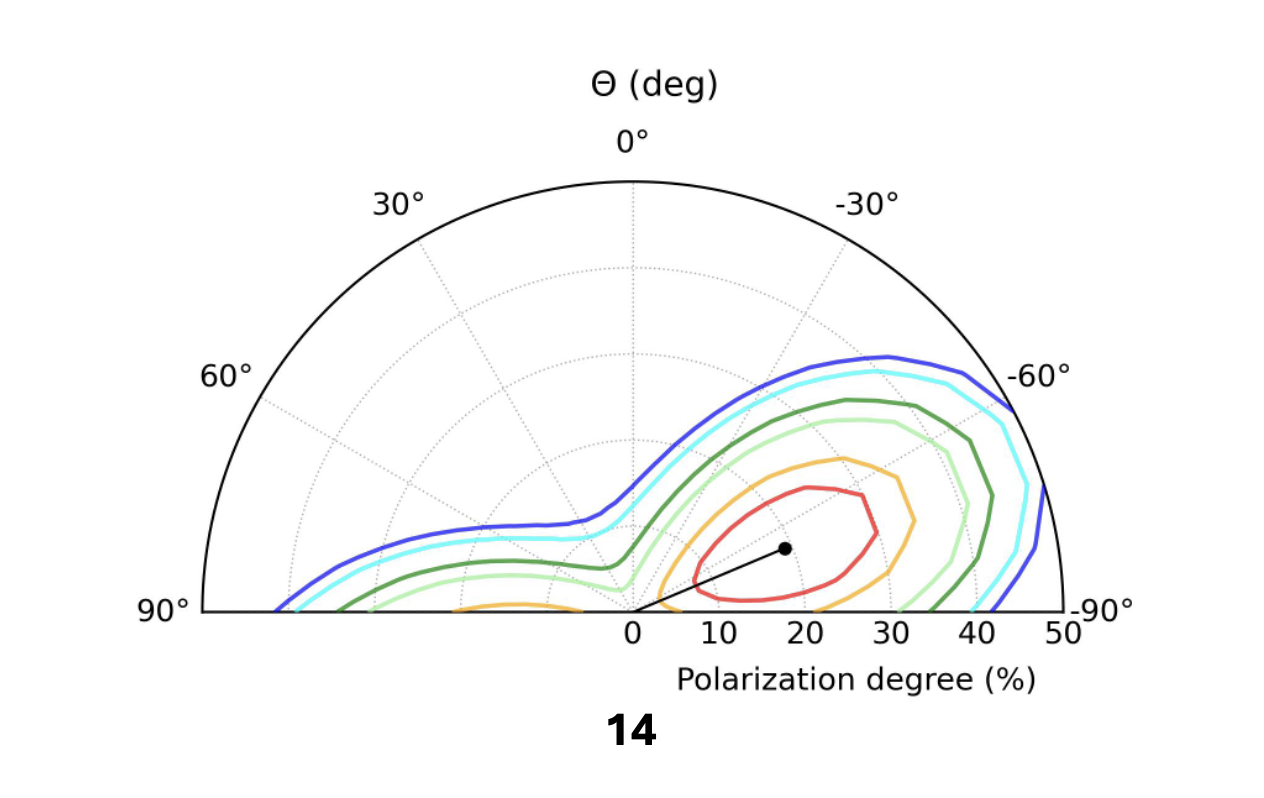}
    \includegraphics[width=0.34\linewidth]{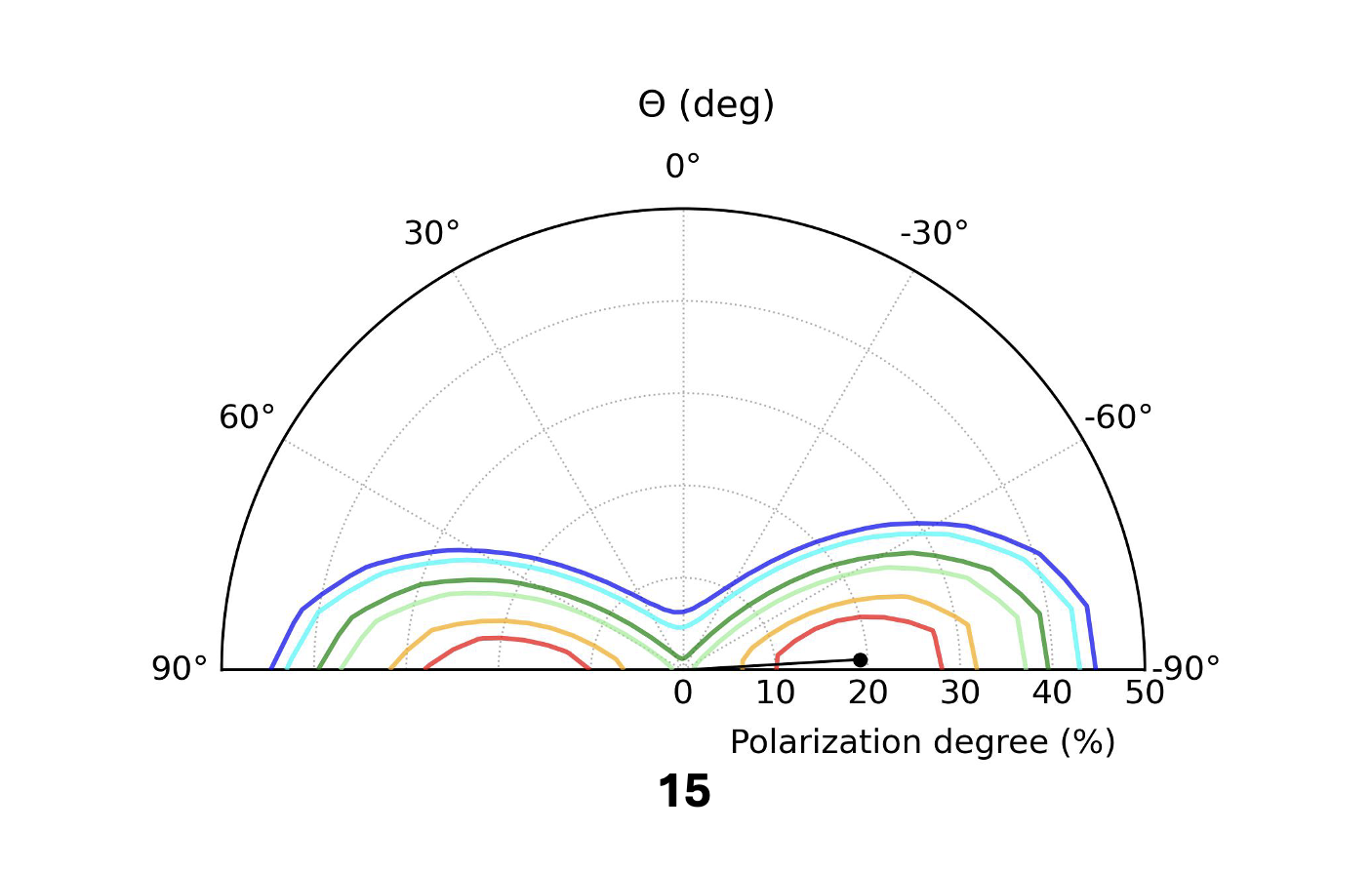}
    \includegraphics[width=0.34\linewidth]{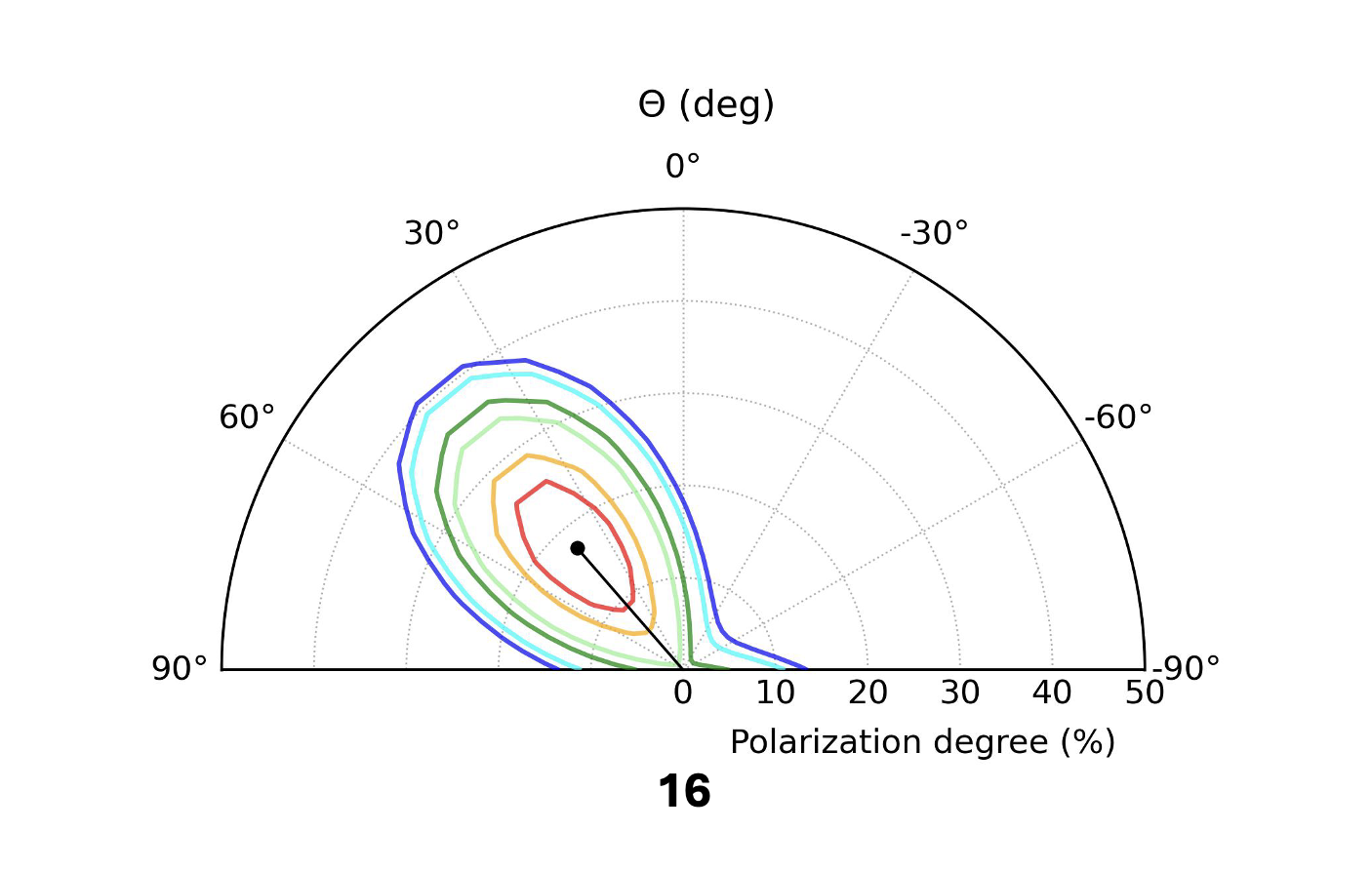}
    \includegraphics[width=0.34\linewidth]{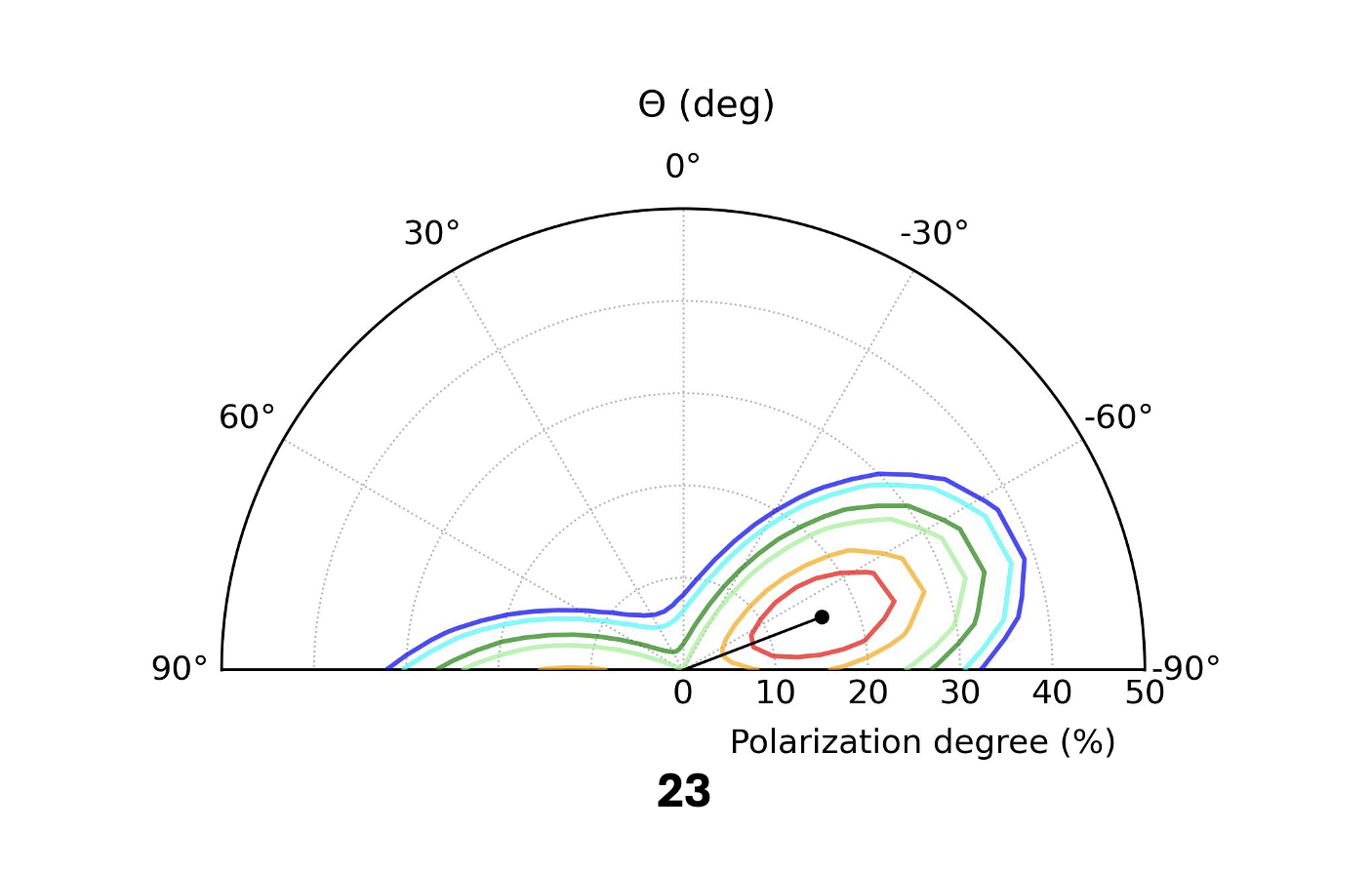}
    \caption{Spectopolarimetric analysis contour plots. Each polar plot shows the PD (\%) and angles in the radial and position angle coordinates, respectively, with the background subtracted. The best-fit PD and PA values are denoted by a black dot. The confidence levels of \(68.27\%\), \(90\%\), \(99\%\), \(99.73\%\), \(99.97\%\), and \(99.99\%\) (based on \(\chi^2\) with 2 degrees of freedom) are represented by the colors red, orange, light green, green, cyan, and blue, respectively.}
    \label{contours}
\end{figure}

\begin{figure}[!h]
    \includegraphics[width=0.32\linewidth]{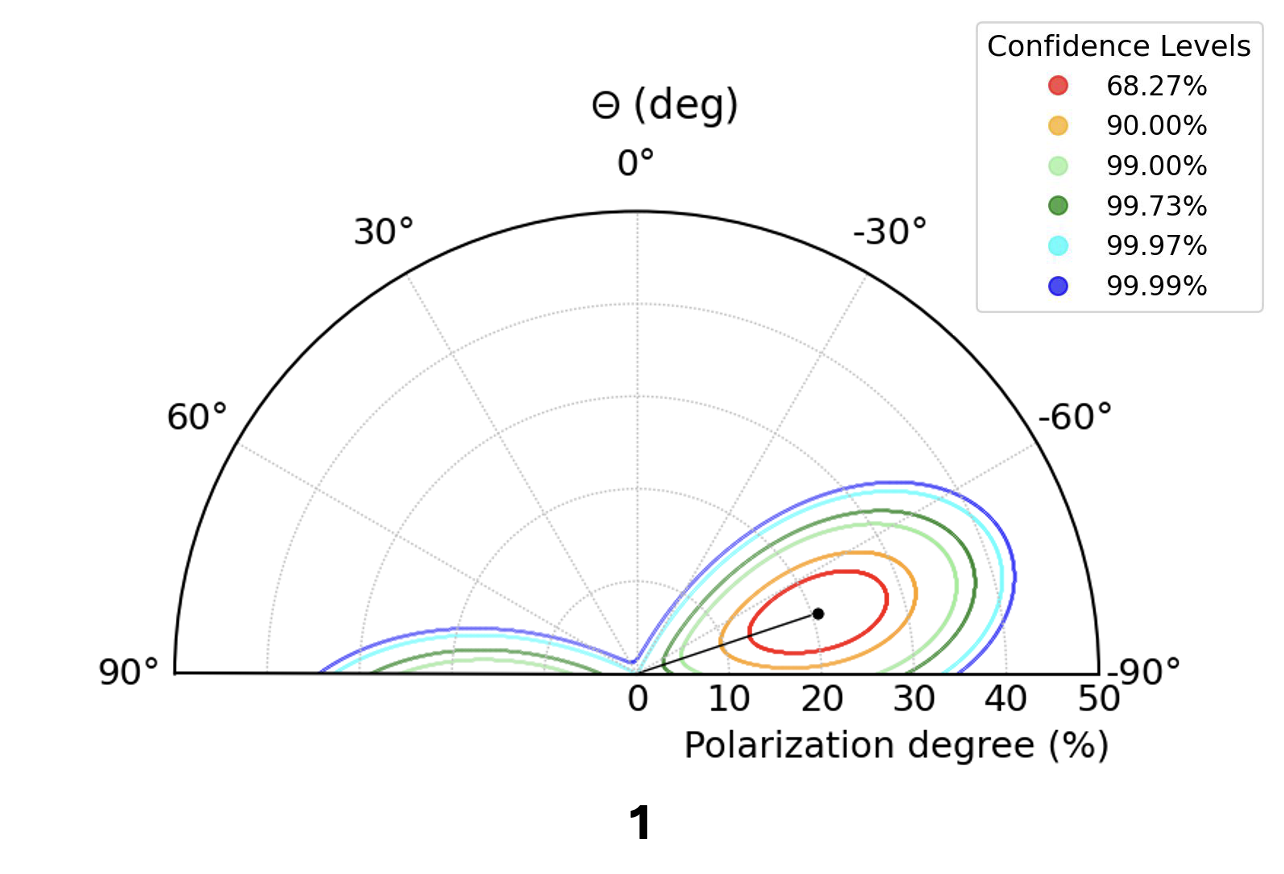}
    \includegraphics[width=0.32\linewidth]{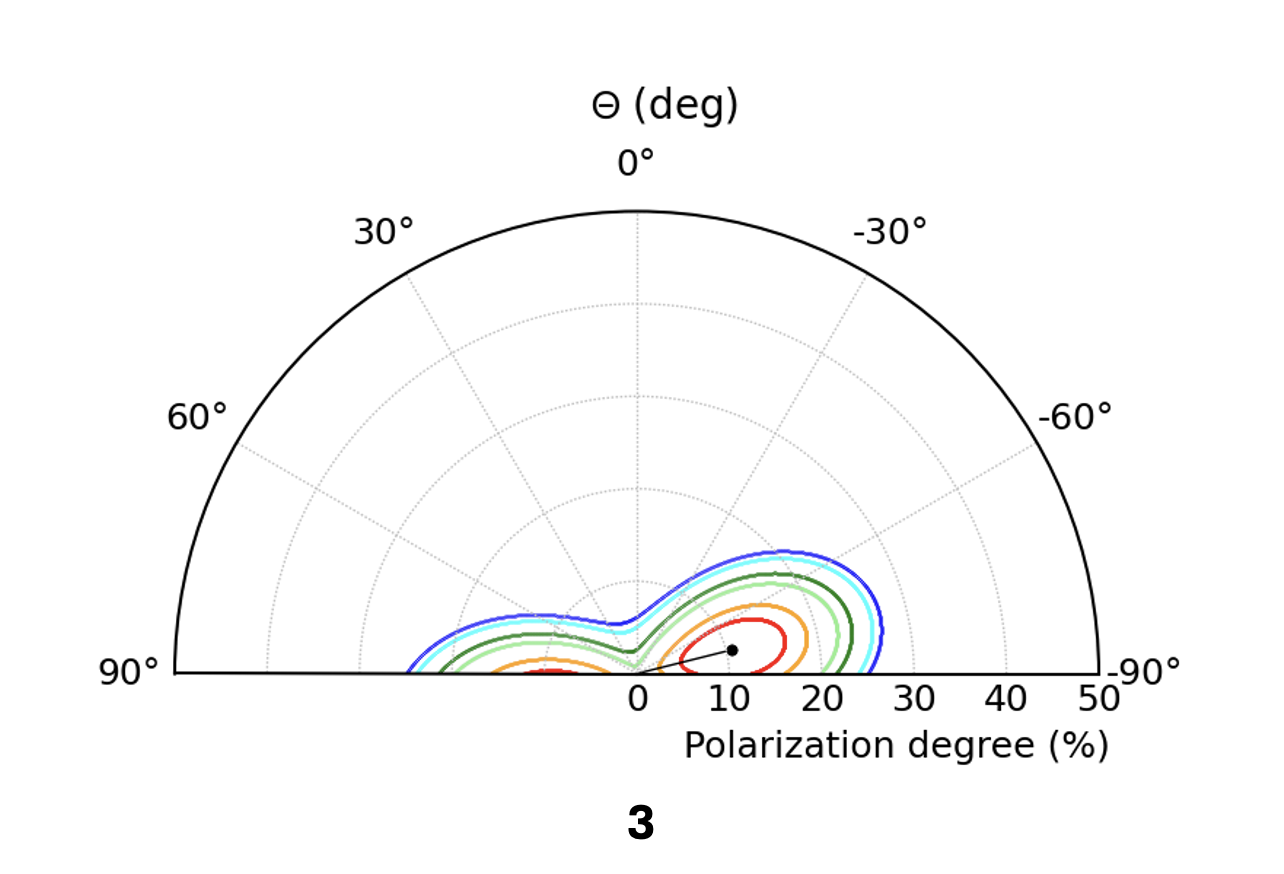}
    \includegraphics[width=0.32\linewidth]{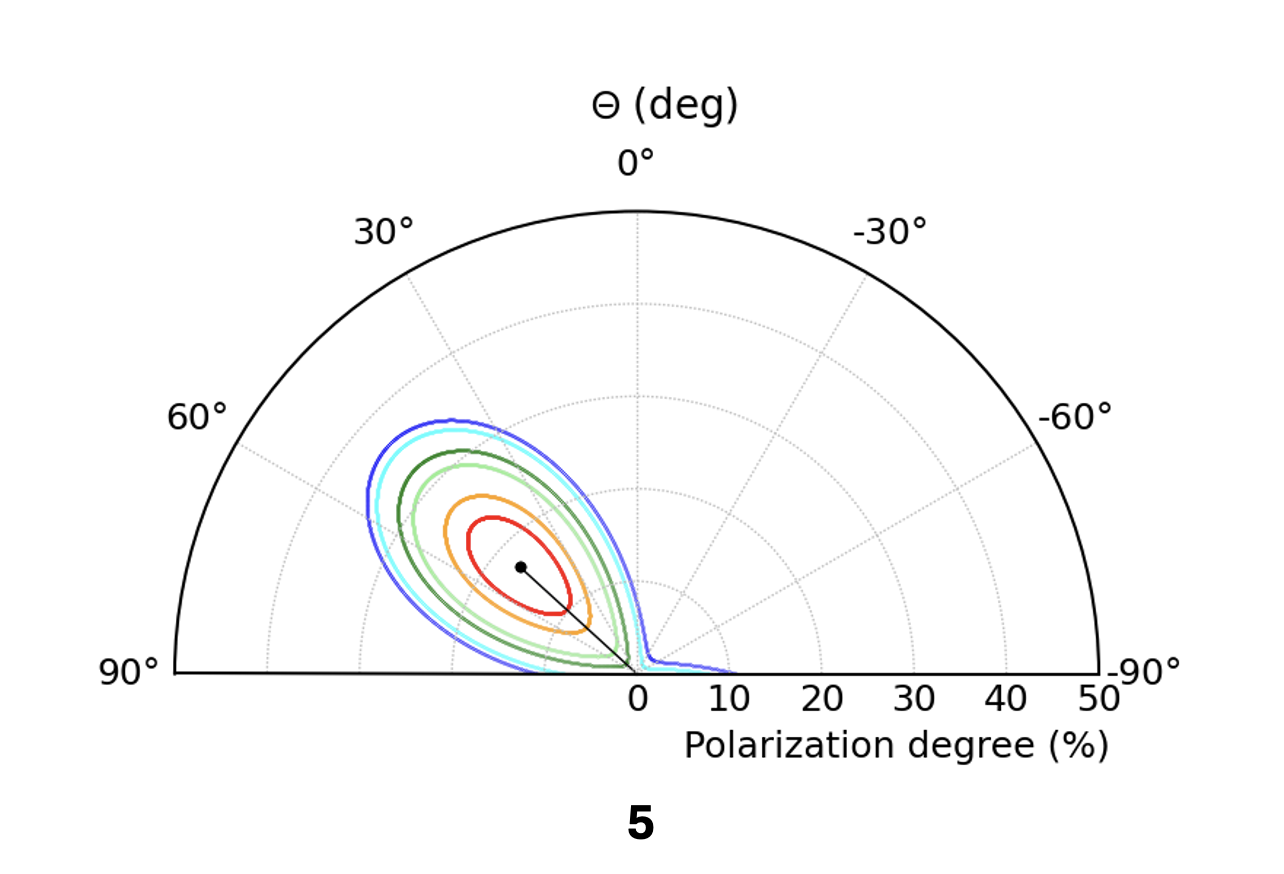}
    \includegraphics[width=0.32\linewidth]{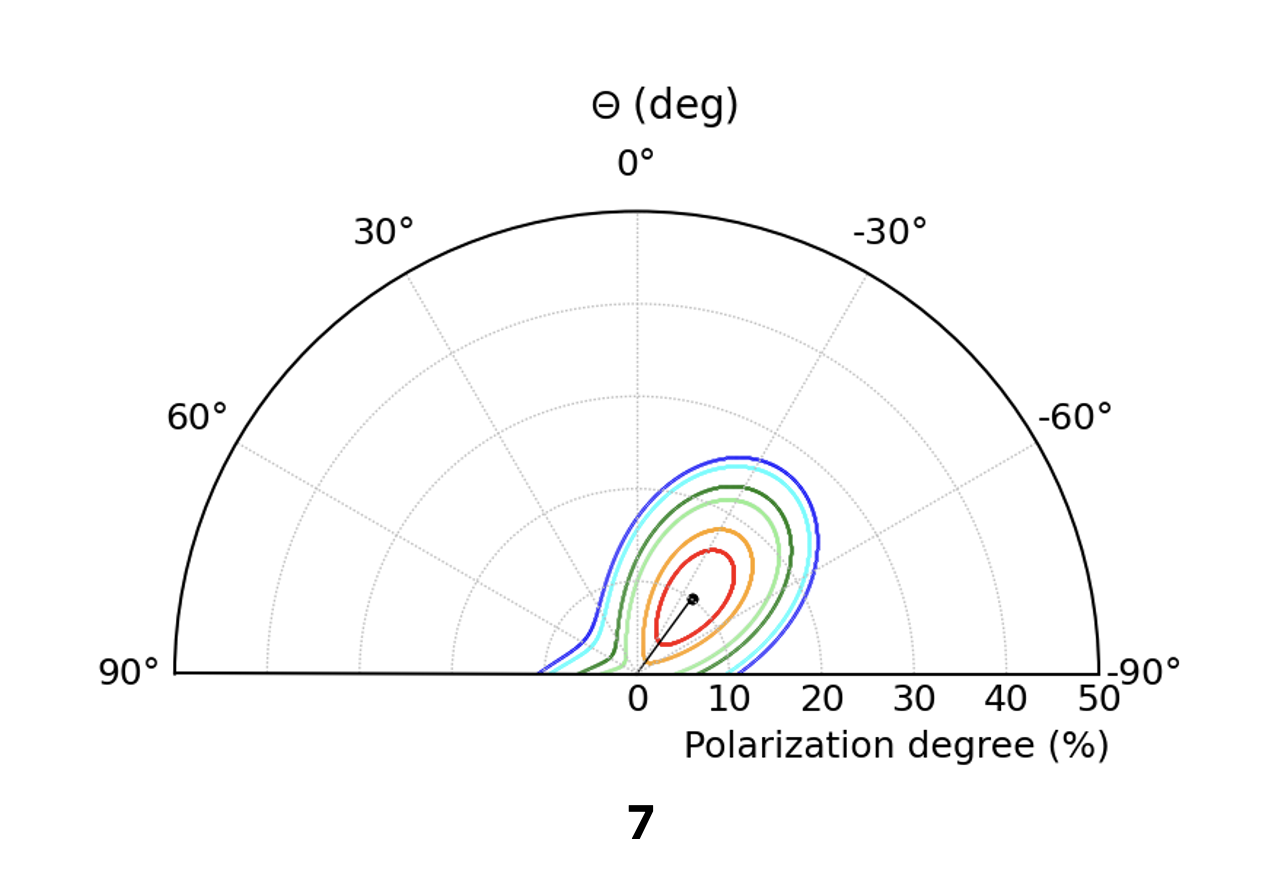}
    \includegraphics[width=0.32\linewidth]{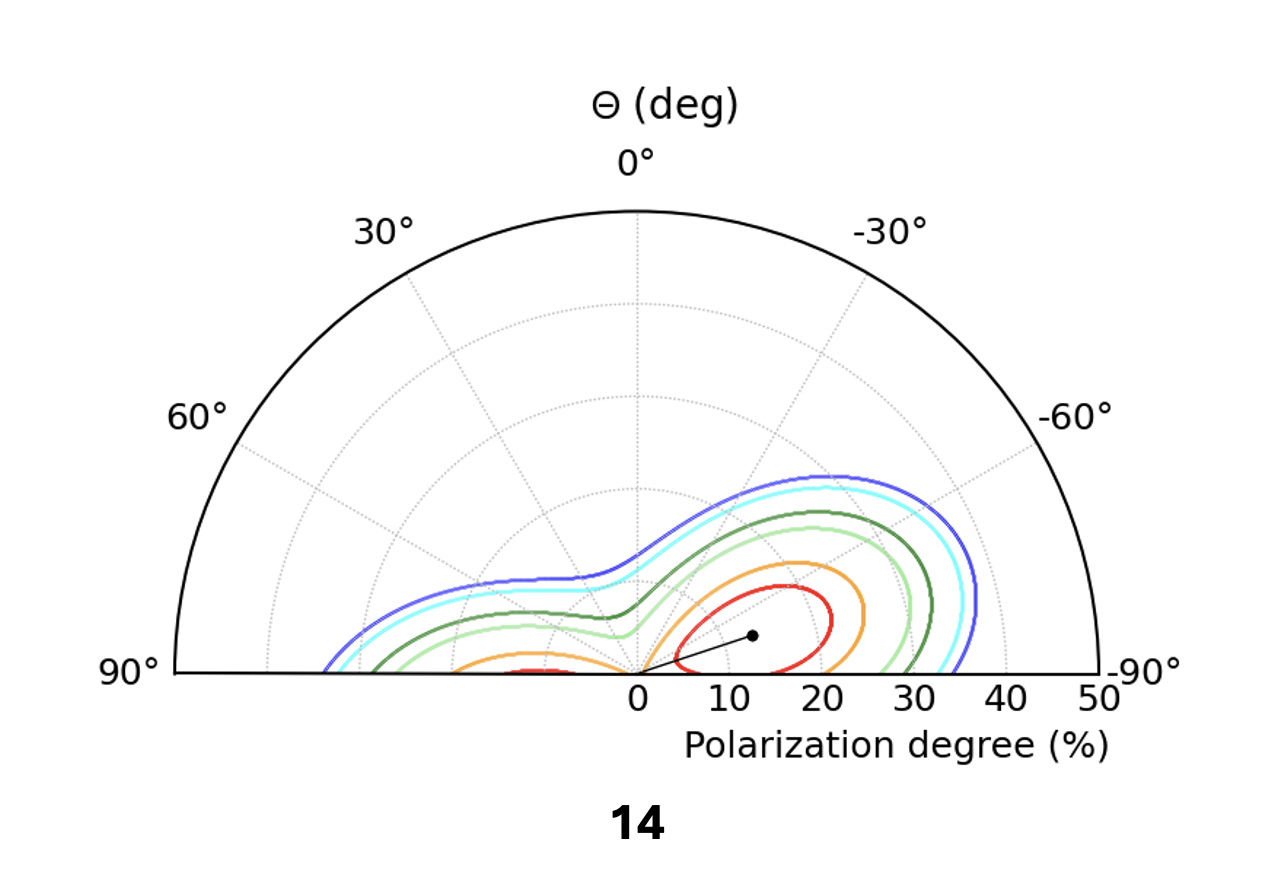}
    \includegraphics[width=0.32\linewidth]{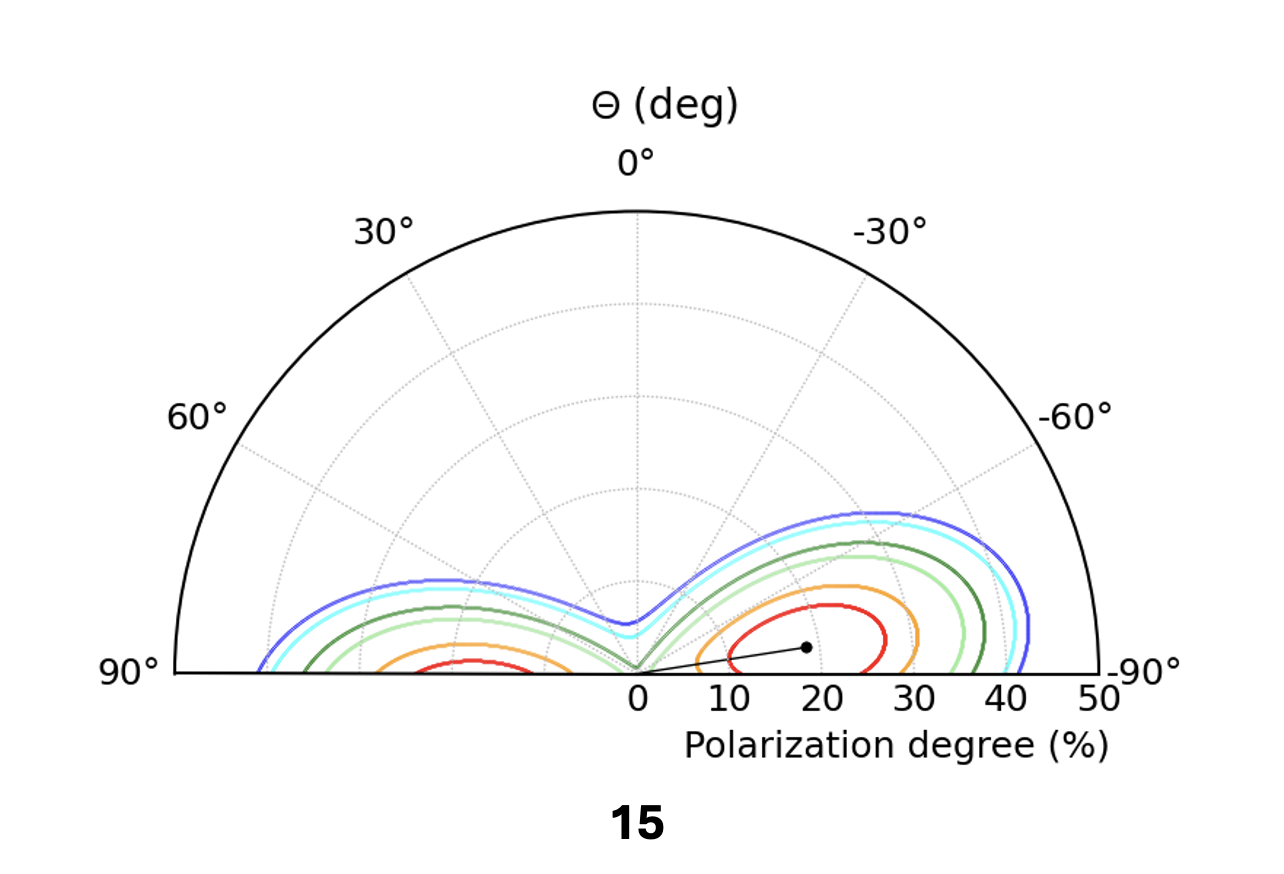}
    \includegraphics[width=0.32\linewidth]{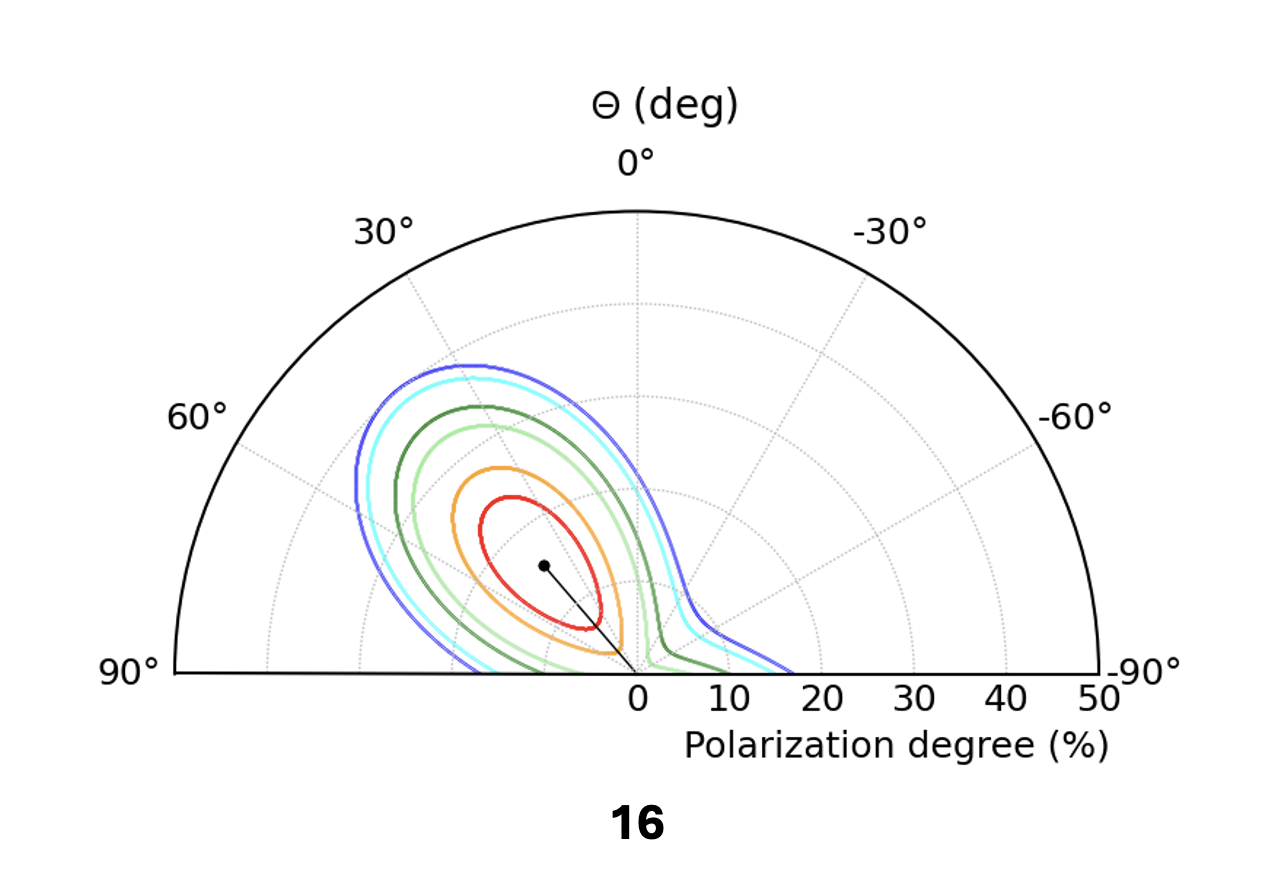}
    \includegraphics[width=0.32\linewidth]{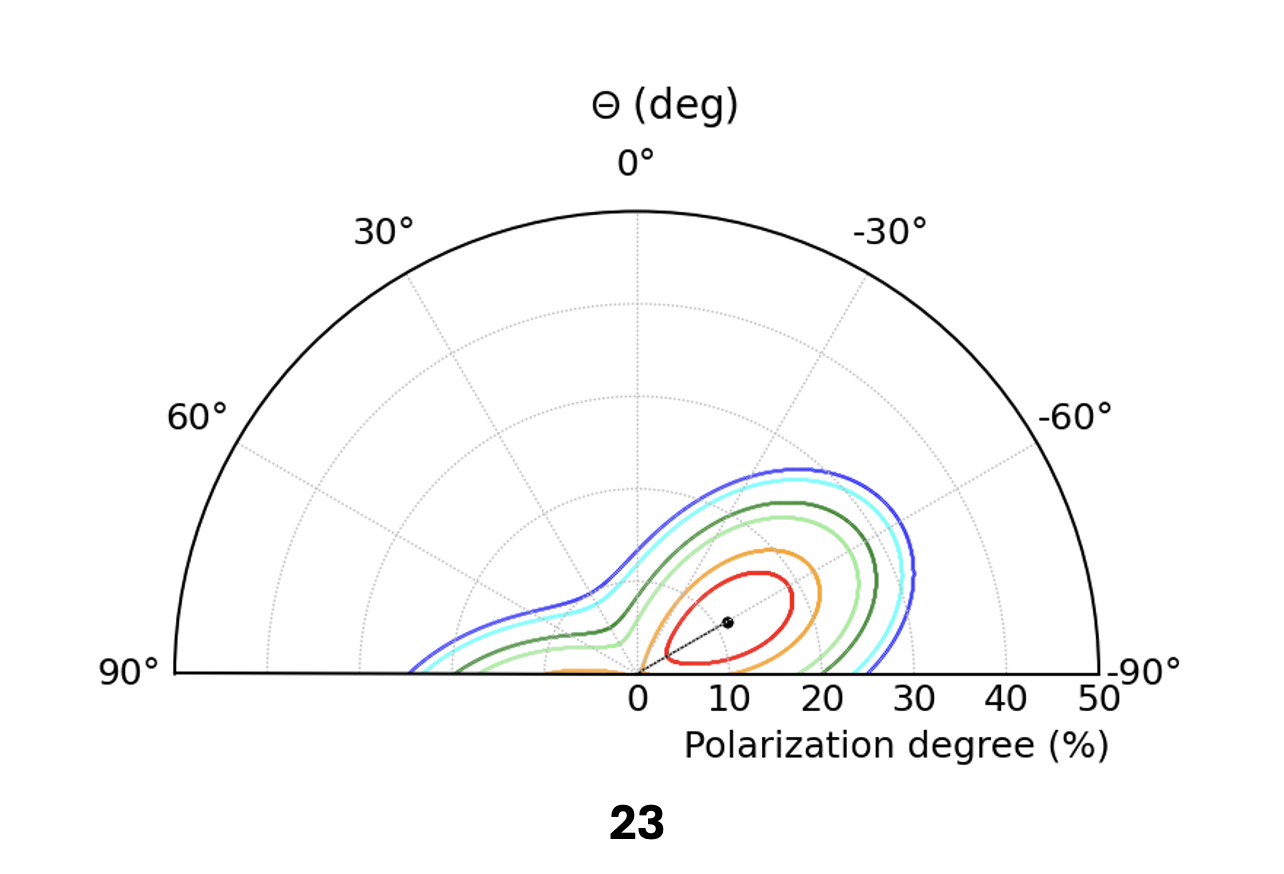}
    \caption{\texttt{IXPEOBSSIM} analysis contour plots. Each polar plot shows the PD (\%) and angles in the radial and position angle coordinates, respectively, with the background subtracted. The PD and PA values obtained by the \texttt{PCUBE} algorithm are denoted by a black dot. The confidence levels of \(68.27\%\), \(90\%\), \(99\%\), \(99.73\%\), \(99.97\%\), and \(99.99\%\) (based on \(\chi^2\) with 2 degrees of freedom) are represented by the colors red, orange, light green, green, cyan, and blue, respectively.}
    \label{contours_2}
\end{figure}

Such droplet-like contours are displayed in Fig. \ref{contours} (for the spectropolarimetric analysis) and Fig. \ref{contours_2} (for the \texttt{IXPEOBSSIM} analysis). They illustrate CLs of 68.27\%, 90\%, 99\%, 99.73\%, 99.97\%, and 99.99\%. The best-fit values for PD and PA are marked with a black dot. Since the exploration of the parameter space involves two variables (PD and PA), the CLs are evaluated using a $\chi^2$ distribution with two degrees of freedom (two-dof). This leads to a lower CL for a fixed $\Delta \chi^2$ compared to the single-parameter case. For example, a CL of 99.90\% corresponds to $\Delta \chi^2_1 = 10.83$ for one degree of freedom and  $\Delta \chi^2_2 = 13.82$ for two degrees of freedom. Conversely, a fixed $\Delta \chi^2 =11.83$ corresponds to a CL of 99.942\% in the one-dof case but drops to 99.73\% in the two-dof scenario.

As a result, fewer regions reach the threshold for significant polarization detection when adopting the two-dof approach. Specifically, regions 4, 11, and 19 fall below the 2$\sigma$ threshold and are therefore excluded from the polarization map shown in Fig. \ref{PD_PA_2}. Nevertheless, many regions still exhibit detections above the 2$\sigma$ threshold, confirming the robustness of the results. Moreover, the analogous analysis performed by \citet{Vink2022} using \texttt{IXPEOBSSIM} also identified fewer regions with significance above 2$\sigma$, with the position of the detected regions on the shell being consistent with those shown in Fig. \ref{PD_PA_2}.

In Fig. \ref{PD_PA_2}, we also show the PD and PA map obtained from the spectral fitting (left panel) and the map of $\chi^2$ values for the polarization signal (right panel) considering the two-dof approach. The latter is set up from \texttt{PMAPCUBE} files generated using \texttt{xpbin}, within the 3–6 keV band, and smoothed via a Welch filter with kernel size of 11 pixels \citep{Prokhorov2024}. This approach makes the results not depend on the size of the pixel. The map of $\chi^2$ values reveals numerous hotspots corresponding to the regions analyzed by spectral analysis. The analysis includes a total of 1109 pixels, resulting in an effective number of trials equal to 9.16. The peak value \( \chi^2 = 23.776 \), which corresponds to a significance level of 99.99931\%, is within our region 1. The post-trials significance is 99.9937\%.\\
Overall, the two-dof analysis aligns fairly well with the standard one with only a few regions falling below the $2\sigma$ threshold and does not significatly affect the results of our analysis.

\section{Estimation of the synchrotron fraction}
Following the method outlined by \citet{Ferrazzoli2023}, we simulate two 5 Ms-long IXPE observations using the \texttt{xpobssim} Monte Carlo simulation tool from \texttt{IXPEOBSSIM}: one accounting for the total emission and the other for the non-thermal emission only. The spectral model is based on \citet{Helder2008}, and we use two Cas A images to derive the remnant’s morphology in the energy bands 1.5-3.0 keV and 4.0-6.0 keV. Given the absence of emission lines in the 4.0-6.0 keV band, we associate this range with the non-thermal emission, while the 1.5-3.0 keV band is attributed to the thermal component.
Using the ratio between the simulated non-thermal and total intensities in the 3.0-6.0 keV band, we estimate the synchrotron fraction for each region of interest, selected from the simulated data with \texttt{xpselect}, reported in Table \ref{sf}. We then divide the background-subtracted polarization, obtained using the \texttt{PCUBE} algorithm in \texttt{xpbin}, by the estimated synchrotron fraction in each region to calculate the corrected polarization degree, PD$_{\text{corr}}$.
\begin{table} [h!]
    \centering
    \tabcolsep=10pt
    \begin{tabular}{c|c}
       \hline
       \hline
       Regions  & Synchrotron fraction ($\%$) \\
       \hline
        1 & 89.48 \\
        3 & 85.90 \\
        4 & 88.38 \\
        5 & 89.39 \\
        7 & 87.69 \\
        11 & 86.47 \\
        14 & 91.01 \\
        15 & 91.13 \\
        16 & 91.52 \\
        19 & 84.65 \\
        23 & 90.42 \\
        \hline
        \hline
    \end{tabular}
    \caption{Fraction of the emission, expressed in (\%), due to non-thermal component in the 3-6 keV energy band for regions showing X-ray polarization at least at the 2$\sigma$ confidence level.}
    \label{sf}
\end{table}

\end{appendices}


\bibliography{bibliography}{}

\begin{thebibliography}{}
\expandafter\ifx\csname natexlab\endcsname\relax\def\natexlab#1{#1}\fi
\providecommand{\url}[1]{\href{#1}{#1}}
\providecommand{\dodoi}[1]{doi:~\href{http://doi.org/#1}{\nolinkurl{#1}}}
\providecommand{\doeprint}[1]{\href{http://ascl.net/#1}{\nolinkurl{http://ascl.net/#1}}}
\providecommand{\doarXiv}[1]{\href{https://arxiv.org/abs/#1}{\nolinkurl{https://arxiv.org/abs/#1}}}

\bibitem[{Amato(2014)}]{Amato2014}
Amato, E. 2014, International Journal of Modern Physics D, 13, 44pp, \dodoi{10.1142/S0218271814300134}

\bibitem[{{Anderson} {et~al.}(1995){Anderson}, {Keohane}, \& {Rudnick}}]{Anderson1995}
{Anderson}, M.~C., {Keohane}, J.~W., \& {Rudnick}, L. 1995, ApJ, 441:300, \dodoi{10.1086/175356}

\bibitem[{{Arnaud}(1996)}]{Arnaud1996}
{Arnaud}, K.~A. 1996, Astronomical Data Analysis Software and Systems V, A.S.P. Conference Series, 101:17

\bibitem[{{Baldini} {et~al.}(2022){Baldini}, {Bucciantini}, {Di Lalla}, { Ehlert}, {Manfreda}, { Negro}, {Omodei}, \& et~al.}]{Baldini2022}
{Baldini}, L., {Bucciantini}, N., {Di Lalla}, N., {et~al.} 2022, SoftwareX, 19, \dodoi{10.1016/j.softx.2022.101194}

\bibitem[{{Bandiera} \& {Petruk}(2016)}]{Bandiera2016}
{Bandiera}, R., \& {Petruk}, O. 2016, MNRAS, 459, 178–198, \dodoi{10.1093/mnras/stw551}

\bibitem[{{Bandiera} \& {Petruk}(2024)}]{Bandiera2024}
---. 2024, A\&A, 689, 14, \dodoi{https://doi.org/10.1051/0004-6361/202450103}

\bibitem[{{Borkowski} {et~al.}(2001){Borkowski}, {Lyerly}, \& {Reynolds}}]{Borkowski2001}
{Borkowski}, K.~J., {Lyerly}, W.~J., \& {Reynolds}, S.~P. 2001, ApJ, 548:820-835, \dodoi{10.1086/319011}

\bibitem[{{Braun} {et~al.}(1987){Braun}, {Gull}, \& {Perley}}]{Braun1987}
{Braun}, R., {Gull}, S.~F., \& {Perley}, R.~A. 1987, Nature, 327, 395, \dodoi{10.1038/327395a0}

\bibitem[{{Bykov} {et~al.}(2024){Bykov}, {Osipov}, {Uvarov}, {Ellison}, \& {Slane}}]{Bykov2024}
{Bykov}, A.~M., {Osipov}, S.~M., {Uvarov}, Y.~A., {Ellison}, D.~C., \& {Slane}, P. 2024, Physical Review D, 110:023041, \dodoi{10.1103/PhysRevD.110.023041}

\bibitem[{{Bykov} {et~al.}(2009){Bykov}, {Uvarov}, {Bloemen}, {den Herder}, \& {Kaastra}}]{Bykov2009}
{Bykov}, A.~M., {Uvarov}, Y.~A., {Bloemen}, J. B. G.~M., {den Herder}, J.~W., \& {Kaastra}, J.~S. 2009, Monthly Notices of the Royal Astronomical Society, 399:3, 1119–1125, \dodoi{https://doi.org/10.1111/j.1365-2966.2009.15348.x}

\bibitem[{{Di Marco} {et~al.}(2023){Di Marco}, {Soffitta}, {Costa}, {Ferrazzoli}, {La Monaca}, {Rankin}, \& et~al.}]{DiMarco2023}
{Di Marco}, A., {Soffitta}, P., {Costa}, E., {et~al.} 2023, ApJ, 165:143, 15pp, \dodoi{10.3847/1538-3881/acba0f}

\bibitem[{{Dickel} \& {Milne}(1976)}]{Dickel1976}
{Dickel}, J.~R., \& {Milne}, D.~K. 1976, Australian Journal of Physics, 29(5), 435, \dodoi{https://doi.org/10.1071/PH760435}

\bibitem[{{Dickel} {et~al.}(1991){Dickel}, {van Breugel}, \& {Strom}}]{Dickel1991}
{Dickel}, J.~R., {van Breugel}, W. J.~M., \& {Strom}, R.~G. 1991, AJ, 101, 2151, \dodoi{10.1086/115837}

\bibitem[{{Drury}(1983)}]{Drury1983}
{Drury}, L.~O. 1983, Reports on Progress in Physics, 46, 973, \dodoi{10.1088/0034-4885/46/8/002}

\bibitem[{{Ferrazzoli} {et~al.}(2024){Ferrazzoli}, {Prokhorov}, {Bucciantini}, {Slane}, {Vink}, \& et~al.}]{Ferrazzoli2024}
{Ferrazzoli}, R., {Prokhorov}, D., {Bucciantini}, N., {et~al.} 2024, ApJL, 967:L38, 15pp, \dodoi{10.3847/2041-8213/ad4a68}

\bibitem[{{Ferrazzoli} {et~al.}(2023){Ferrazzoli}, {Slane}, {Prokhorov}, {Zhou}, {Vink}, \& et~al.}]{Ferrazzoli2023}
{Ferrazzoli}, R., {Slane}, P., {Prokhorov}, D., {et~al.} 2023, ApJ, 945:52, 14pp, \dodoi{10.3847/1538-4357/acb496}

\bibitem[{{Ginzburg} \& {Syrovatskii}(1965)}]{GinzburgSyrovatskii1965}
{Ginzburg}, V.~L., \& {Syrovatskii}, S.~I. 1965, Annual Review of Astronomy and Astrophysics, 3, s97, \dodoi{10.1146/annurev.aa.03.090165.001501}

\bibitem[{{Greco} {et~al.}(2023){Greco}, {Vink}, {Ellien}, \& {Ferrigno}}]{Greco2023}
{Greco}, E., {Vink}, J., {Ellien}, A., \& {Ferrigno}, C. 2023, \apj, 956, 116, \dodoi{10.3847/1538-4357/acf567}

\bibitem[{{Helder} \& {Vink}(2008)}]{Helder2008}
{Helder}, E.~A., \& {Vink}, J. 2008, ApJ, 686, 1094, \dodoi{https://doi.org/10.1086/591242}

\bibitem[{{Helder} {et~al.}(2012){Helder}, {Vink}, {Bykov}, {Ohira}, J.C.~{Raymond}, \& R.}]{Helder2012}
{Helder}, E.~A., {Vink}, J., {Bykov}, A., {et~al.} 2012, Space Science Reviews, 173, 369–431, \dodoi{https://doi.org/10.1007/s11214-012-9919-8}

\bibitem[{{Hwang} {et~al.}(2004){Hwang}, {Laming}, {Badenes}, {Berendse}, \& et~al.}]{Hwang2004}
{Hwang}, U., {Laming}, J.~M., {Badenes}, C., {Berendse}, F., \& et~al. 2004, ApJ, 615:L117–L120, \dodoi{10.1086/426186}

\bibitem[{{Kelner} {et~al.}(2013){Kelner}, {Aharonian}, \& {Khangulyan}}]{Kelner2013}
{Kelner}, S.~R., {Aharonian}, F.~A., \& {Khangulyan}, D. 2013, \apj, 774, 61, \dodoi{10.1088/0004-637X/774/1/61}

\bibitem[{{Krause} {et~al.}(2008){Krause}, {Birkmann}, {Usuda}, {Hattori}, {Goto}, {Rieke}, \& {Misselt}}]{Krause2008}
{Krause}, O., {Birkmann}, S.~M., {Usuda}, T., {et~al.} 2008, Science, 320:1195, \dodoi{10.1126/science.1155788}

\bibitem[{{Kundu} \& {Velusamy}(1971)}]{Kundu&Velusamy1971}
{Kundu}, M.~R., \& {Velusamy}, T. 1971, ApJ, 163, 231, \dodoi{10.1086/150761}

\bibitem[{{Lawrence} {et~al.}(1995){Lawrence}, {MacAlpine}, {Uomoto}, {Woodgate}, \& et~al.}]{Lawrence1995}
{Lawrence}, S.~S., {MacAlpine}, G.~M., {Uomoto}, A., {Woodgate}, B.~E., \& et~al. 1995, ApJ, 108:2635, \dodoi{10.1086/117477}

\bibitem[{{Milisavljevic1} \& {Fesen}(2013)}]{Milisavljevic12013}
{Milisavljevic1}, D., \& {Fesen}, R.~A. 2013, ApJ, 772:134, \dodoi{10.1088/0004-637X/772/2/134}

\bibitem[{{Morlino} {et~al.}(2010){Morlino}, {Amato}, {Blasi}, \& {Caprioli}}]{Morlino2010}
{Morlino}, G., {Amato}, E., {Blasi}, P., \& {Caprioli}, D. 2010, Monthly Notices of the Royal Astronomical Society, 405, L21, \dodoi{https://doi.org/10.1111/j.1745-3933.2010.00851.x}

\bibitem[{{Orlando} {et~al.}(2022){Orlando}, {Wongwathanarat}, {Janka}, {Miceli}, {Nagataki}, {Ono}, {Bocchino}, {Vink}, {Milisavljevic}, {Patnaude}, \& {Peres}}]{Orlando2022}
{Orlando}, S., {Wongwathanarat}, A., {Janka}, H.~T., {et~al.} 2022, \aap, 666, A2, \dodoi{10.1051/0004-6361/202243258}

\bibitem[{{Parizot} {et~al.}(2006){Parizot}, {Marcowith}, {Ballet}, \& {Gallant}}]{Parizot2006}
{Parizot}, E., {Marcowith}, A., {Ballet}, J., \& {Gallant}, Y.~A. 2006, Astronomy \& Astrophysics, 453:387–395, \dodoi{10.1051/0004-6361:20064985}

\bibitem[{{Prokhorov} {et~al.}(2024){Prokhorov}, {Yang}, {Vink}, {Slane}, {Costa}, \& et~al.}]{Prokhorov2024}
{Prokhorov}, D.~A., {Yang}, Y., {Vink}, J., {et~al.} 2024, Astronomy \& Astrophysics, \dodoi{https://doi.org/10.1051/0004-6361/202452062}

\bibitem[{{Reed} {et~al.}(1995){Reed}, {Hester}, {Fabian}, \& {Winkler}}]{Reed1995}
{Reed}, J.~E., {Hester}, J.~J., {Fabian}, A.~C., \& {Winkler}, P.~F. 1995, ApJ, 440:706, \dodoi{10.1086/175308}

\bibitem[{{Reynoso} {et~al.}(2013){Reynoso}, {Hughes}, \& {Moffett}}]{Reynoso2013}
{Reynoso}, E.~M., {Hughes}, J.~P., \& {Moffett}, D.~A. 2013, AJ, 145, 104, \dodoi{10.1088/0004-6256/145/4/104}

\bibitem[{{Rosenberg}(1970)}]{Rosenberg1970}
{Rosenberg}, I. 1970, MNRAS, 151, 109, \dodoi{10.1093/mnras/151.1.109}

\bibitem[{{Sakai} {et~al.}(2024){Sakai}, {Yamada}, {Sato}, {Hayakawa}, \& {Kominato}}]{Sakai2024}
{Sakai}, Y., {Yamada}, S., {Sato}, T., {Hayakawa}, R., \& {Kominato}, N. 2024, The Astrophysical Journal, 974:245, 16pp, \dodoi{https://doi.org/10.3847/1538-4357/ad739f}

\bibitem[{{Slane} {et~al.}(1999){Slane}, {Gaensler}, {Dame}, {Hughes}, {Plucinsky}, \& {Green}}]{Slane1999}
{Slane}, P., {Gaensler}, B., {Dame}, T.~M., {et~al.} 1999, ApJ, 525:357-367, \dodoi{10.1086/307893}

\bibitem[{{Soffitta} {et~al.}(2021){Soffitta}, {Baldini}, {Bellazzini}, {Costa}, {Latronico}, \& et~al.}]{Soffitta2021}
{Soffitta}, P., {Baldini}, L., {Bellazzini}, R., {et~al.} 2021, ApJ, 162:208, 18pp, \dodoi{10.3847/1538-3881/ac19b0}

\bibitem[{{Toptygin} \& {Fleishman}(1987)}]{Toptygin1987}
{Toptygin}, I.~N., \& {Fleishman}, G.~D. 1987, \apss, 132, 213, \dodoi{10.1007/BF00641755}

\bibitem[{Vink(2020)}]{Vink2020}
Vink, J. 2020, Physics and Evolution of Supernova Remnants (Astronomy and Astrophysics Library, Springer), \dodoi{10.1007/978-3-030-55231-2}

\bibitem[{{Vink} {et~al.}(2022{\natexlab{a}}){Vink}, {Patnaude}, \& {Castro}}]{Vink2022b}
{Vink}, J., {Patnaude}, D.~J., \& {Castro}, D. 2022{\natexlab{a}}, \apj, 929, 57, \dodoi{10.3847/1538-4357/ac590f}

\bibitem[{{Vink} {et~al.}(2022{\natexlab{b}}){Vink}, {Prokhorov}, {Ferrazzoli}, {Slane}, {Zhou}, \& et~al.}]{Vink2022}
{Vink}, J., {Prokhorov}, D., {Ferrazzoli}, R., {et~al.} 2022{\natexlab{b}}, ApJ, 938:40, 14pp, \dodoi{10.3847/1538-4357/ac8b7b}

\bibitem[{{Weisskopf} {et~al.}(2022){Weisskopf}, {Soffitta}, {Baldini}, {Ramsey}, {O’Dell}, \& et~al.}]{Weisskopf2022}
{Weisskopf}, M.~C., {Soffitta}, P., {Baldini}, L., {et~al.} 2022, J. Astron. Telesc. Instrum. Syst., 8(2), \dodoi{10.1117/1.JATIS.8.2.026002}

\bibitem[{{Wentzel}(1974)}]{Wentzel1974}
{Wentzel}, D.~G. 1974, Annual review of astronomy and astrophysics, 12, 71, \dodoi{10.1146/annurev.aa.12.090174.000443}

\bibitem[{{Wilms} {et~al.}(2000){Wilms}, {Allen}, \& {McCray}}]{Wilms2000}
{Wilms}, J., {Allen}, A., \& {McCray}, R. 2000, ApJ, 542:914-924, \dodoi{10.1086/317016}

\bibitem[{{Zhou} {et~al.}(2023){Zhou}, {Prokhorov}, {Yang}, {Ferrazzoli}, {Slane}, {Vink}, \& et~al.}]{Zhou2023}
{Zhou}, P., {Prokhorov}, D., {Yang}, Y.~J., {et~al.} 2023, ApJ, 957:55, 12pp, \dodoi{10.3847/1538-4357/acf3e6}

\end{thebibliography}
\bibliographystyle{aasjournal}

\end{document}